\documentclass[usegraphicx,useAMS,usenatbib]{mn2e}

\usepackage{graphicx}
\usepackage{times}
\usepackage{amssymb}
\usepackage{amsmath}
\usepackage{aas_macros}

\def\nW{$\mathrm{nWm^{-2}sr^{-1}}$}
\def\mic{$\mu {\rm m}$}
\def\gsim{\gtrsim}
\def\lsim{\lesssim}

\begin{document}

\title{ On the Physical Requirements for a Pre-Reionization Origin of the Unresolved Near-Infrared Background }

\title{ On the Physical Requirements for a Pre-Reionization Origin of the Unresolved Near-Infrared Background }
\author[K. Helgason, M. Ricotti, A. Kashlinsky, V. Bromm]
{K. Helgason\thanks{E-mail: kari@mpa-garching.mpg.de }$^1$, M. Ricotti$^2$, A. Kashlinsky$^{3,4}$, V. Bromm$^5$ \\
  $^1$Max Planck Institute for Astrophysics, Karl-Schwarzschild-Str. 1, 85748 Garching, Germany  \\
$^2$Department of Astronomy, University of Maryland, College Park, MD 20742 \\
$^3$Observational Cosmology Laboratory, Code 665, NASA Goddard Space Flight Center, Greenbelt MD 20771 \\
$^4$SSAI, Lanham, MD 20706 \\
$^5$Department of Astronomy, University of Texas, Austin, TX 78712}

\maketitle

\begin{abstract}

The study of the Cosmic Near-Infrared Background (CIB) light after subtraction of resolved sources can push the limits of current observations and yield information on galaxies and quasars in the early universe. Spatial fluctuations of the CIB exhibit a clustering excess at angular scales $\sim 1^\circ$ whose origin has not been conclusively identified, but disentangling the relative contribution from low- and high-redshift sources is not trivial. We explore the likelihood that this signal is dominated by emission from galaxies and accreting black holes in the early Universe. We find that, the measured fluctuation signal is too large to be produced by galaxies at redshifts $z>8$, which only contribute $\sim 0.01-0.05$ \nW\ to the CIB. Additionally, if the first small mass galaxies have a normal IMF, the light of their ageing stars (fossils) integrated over cosmic time contributes a comparable amount to the CIB as their pre-reionization progenitors. In order to produce the observed level of CIB fluctuation without violating constraints from galaxy counts and the electron optical depth of the IGM, minihalos at $z>12$ must form preferably top-heavy stars with efficiency $f_\star \gsim 0.1$ and at the same time maintain a very low escape fraction of ionizing radiation, $f_{\rm esc}<0.1\%$. If instead the CIB fluctuations are produced by high-$z$ black holes, one requires vigorous accretion in the early universe reaching $\rho_{\rm acc} \gsim 10^5M_\odot{\rm Mpc^{-3}}$ by $z\simeq 10$. This growth must stop by $z \sim 6$ and be significantly obscured not to overproduce the soft cosmic X-ray background (CXB) and its observed coherence with the CIB. We therefore find the range of suitable high-$z$ explanations to be narrow, but could possibly be widened by including additional physics and evolution at those epochs.

\end{abstract}


\section{ Introduction }

The Cosmic Near Infrared Background (CIB) contains radiation that has been built up throughout the cosmic history, including highly redshifted emission from the pre-reionization era \citep[see e.g. review by ][]{Kashreview}. Measured brightness fluctuations in deep source-subtracted CIB maps have established the existence of an unresolved CIB component in addition to resolved point sources \citep{KAMM1,KAMM2,Thompson07a,Thompson07b,Matsumoto11,Kashlinsky12,Cooray12b,Seo15}. Whereas the CIB flux associated with this component cannot be directly determined from such measurements, the required levels are nevertheless theoretically deduced \citep{KAMM1} to lie well below earlier studies claiming high CIB \citep{DwekArendt98,Matsumoto05,Tsumura13,Matsumoto15} in excess of integrated galaxy counts \citep{MadauPozzetti00,Fazio04,Keenan10a,Ashby13}; and at the same time consistent with indirect measurements from $\gamma$-ray blazars \citep{MazinRaue07,Meyer12,Ackermann12,HESS13,Biteau15}.

The unresolved CIB fluctuations measured with {\it Spitzer}/IRAC \citep{KAMM1,KAMM2,Kashlinsky12,Cooray12b}, {\it HST}/NICMOS \citep{Thompson07a,Thompson07b} and {\it AKARI}/IRC \citep{Matsumoto11,Seo15}, confirm a mutually consistent isotropic signal above the noise extending out to $\sim 1^\circ$ scales. The observed properties can be listed as follows (see \citet{Kashlinsky15} for detailed discussion): 
1) the signal is inconsistent with local foregrounds such as the Zodiacal Light and Galactic cirrus \citep{KAMM1,Arendt10,Matsumoto11,Kashlinsky12}. Its extragalactic nature is further supported by its isotropy (now measured in several different parts of the sky, \citet{Kashlinsky12});
2) the signal is also inconsistent with the contribution from known galaxy populations at $z<6$ extrapolated to faint luminosities \citep{Helgason12};
3) the amplitude of the fluctuations increases towards shorter wavelengths showing a blue spectrum \citep{Matsumoto11,Seo15};
4) 4) there is no evidence yet for the fluctuations correlating significantly with the mask or the outer parts of removed sources \citep{KAMM1,Arendt10}; 
5) the large-scale clustering component does not yet appear to start decreasing as the small-scale shot noise power is lowered;
6) the clustering component of the fluctuations shows no correlation with faint HST/ACS source maps down to $m_{\rm AB} \sim 28$ \citep{KAMM4},
7) the unresolved CIB fluctuations at 3.6\mic\ are coherent with the Cosmic Far-Infrared Background (200,350,500\mic) which can largely be explained by unresolved galaxies at low-$z$ and their extended emission \citep{Thacker14},
8) the source-subtracted CIB fluctuations are coherent with the unresolved soft Cosmic X-ray Background (not detected in harder X-ray bands
of $>2$keV) \citep{Cappelluti13} which can partly be accounted for by unresolved AGN and X-ray galaxies at low-$z$ \citep{Helgason14}.
A recent measurement by \citet{Zemcov14} finds the blue spectrum continuing to shorter wavelengths but appears to be in conflict with earlier {\it HST}/NICMOS and 2MASS fluctuation studies at the same wavelengths from 2MASS \citep{Kashlinsky02} and {\it HST}/NICMOS \citep{Thompson07a}.

It was suggested that the era of the first stars could have left a measurable imprint in the CIB, both its mean level \citep{Santos02} and its anisotropies \citep{Cooray04,Kashlinsky04}. This was followed by more detailed studies on the nature of these populations \citep{Salvaterra06a,Salvaterra06b,Fernandez06,KAMM3,Fernandez10}. This was motivated by the expectation that the first objects were i) individually bright with a short epoch of energy release, ii) highly biased as they form out of rare density peaks, and iii) radiate strongly in UV/blue being redshifted into today's near-IR part of the spectrum \citep[reviewed in ][]{Bromm13}. Despite being consistent with all observed properties however, the high-$z$ origin of the source-subtracted CIB signal continues to be debated. This is in part because of the lack of a robust redshift determination of the signal and because recent models of early galaxy populations $z\gsim 6$ have failed to produce sufficient CIB fluctuation power \citep{Fernandez12,Cooray12a,Yue13a}. This has motivated alternative hypotheses for their origin, such as in a diffuse intrahalo light at low/intermediate-$z$ \citep{Cooray12b,Zemcov14}. At the same time, it was suggested that accretion by direct collapse black holes (DCBH) at high-$z$ can provide an explanation which also fully accounts for the CIB$\times$CXB signal \citep{Yue13b}.

We evaluate the physical requirements for a pre-reionization origin of the CIB fluctuations based on the latest observational insights. We carefully quantify the clustering excess in Section \ref{sec:excess}. In Section  \ref{sec:stars}  we use an analytic approach to model the CIB contributions from early stellar populations, both metal-free and metal-enriched. In Section \ref{sec:bhs} we consider accreting black holes and associated gaseous emission. We derive the required star formation and accretion rates and discuss their implications for the CXB. In this paper we use the standard $\Lambda$CDM cosmology with parameters ($h$, $\Omega_{\rm m}$, $\Omega_{\rm \Lambda}$, $n_s$, $\sigma_8$)=(0.678,0.308,0.692,0.968,0.829) \citep{PlanckCosmology}. All magnitudes are in the AB system \citep{OkeGunn83}.

\section{ Quantifying the CIB Fluctuation Excess } \label{sec:excess}

\begin{figure*}
\begin{center}
      \includegraphics[width=0.98\textwidth]{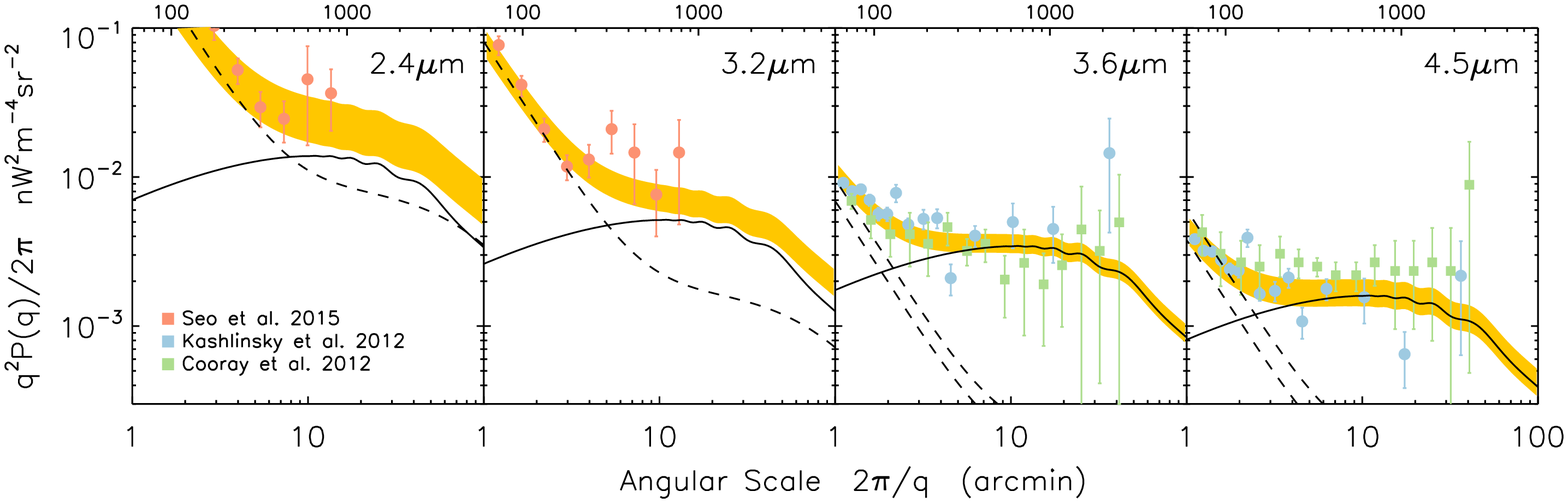}
      \caption{ The latest measurements of the mean-square CIB fluctuations in the 2-5\mic\ range. AKARI/IRC measurements are shown as red circles \citep{Seo15} whereas Spitzer/IRAC are shown in blue circles \citep{Kashlinsky12} and green squares \citep{Cooray12b}. The solid curve shows the best-fit $\Lambda$CDM power spectrum projected from $z=50$ to $z=10$ whereas the dashed lines correspond to the unresolved galaxy populations from \citet{Helgason12} fitted to each individual measurement (see text). The fit gives ${\rm \chi^2/dof=1.62}$. The upper horizontal axis shows the angular scale in arcseconds. The large scale averaged fluctuations in each band are $\delta F(\langle 3^\prime-30^\prime \rangle)= (0.11,0.068,0.056,0.037)^{+( 0.044,0.013,0.006,0.01)}_{-( 0.037,0.012,0.006,0.009)}$\nW at (2.4,3.2,3.6,4.5)\mic. }
      \label{fig_measurements}
\end{center}
\end{figure*}

Brightness fluctuations in the CIB can be written $\delta F({\bf x}) = F({\bf x}) - \langle F \rangle$ where $F({\bf x})$ is the sky brightness at the 2-D coordinate ${\bf x}$ and $\langle F \rangle$ is the mean isotropic flux. We describe the fluctuation field in terms of the power spectrum as a function of the angular wavenumber $q$, defined as $P(q)=\langle |\delta_{\bf q}|^2\rangle$, where $\delta_{\bf q}$ is the 2-D Fourier transform of the CIB fluctuation field, $\delta F({\bf x})$. On the angular scale $\theta=2\pi /q$, the root-mean-square fluctuation in the CIB can be written as, $\delta F_{\bf \theta} \equiv \langle \delta F^2({\bf x})\rangle ^{1/2} \simeq [q^2P(q)/2\pi ]^{1/2}$. The fluctuations are therefore determined by the flux of the underlying sources and how they cluster on the sky
\begin{equation} \label{eqn_fracfluc}
  \delta F_\theta = F_{\rm CIB}\Delta_\theta
\end{equation}
where the information on the clustering as a function of angular scale is contained within the fractional fluctuations $\Delta_\theta$. At high-$z$, the sources trace increasingly rare peaks in the density field and $\Delta_{\rm cl}$ can become $\simeq 0.1$ at arcminute scales. Current fluctuation measurements exhibit a signal above the noise which flattens to an approximately constant rms value towards large angular scales (see Figure \ref{fig_measurements}). In this paper we refer to the {\it clustering} as the average of this large scale value $\Delta_{\rm cl}  \equiv \langle {\Delta}(3^\prime-30^\prime)\rangle$. When defining our CIB fluctuation excess, we focus on NIR fluctuations at $2<\lambda<5$\mic\ from {\it AKARI}/IRC and {\it Spizter}/IRAC. This is where current measurements show a mutually consistent large scale signal which at the same time are deep enough such that the clustering component is not sensitive to further decreasing the shot noise from remaining galaxies.

In order to get a single representative value for the measured large scale CIB fluctuations from high-$z$, we consider a simple three-parameter model consisting of the sum of fluctuations from unresolved galaxy populations at the measurement detection threshold \citep{Helgason12} and $\Lambda$CDM power spectrum\footnote{The power spectrum is normalized at the 10$^\prime$ peak such that $\delta F_{\rm 3.6\mu m} = 0.043$\nW.} \citep{EisensteinHu98}, projected to $z=10$:
\begin{equation} \label{eqn_fit}
\delta F_\theta^2(\lambda)~=~ {\rm a}\cdot\delta F_{\rm \Lambda CDM}^2\left(\frac{\lambda}{3.6{\rm \mu m}}\right)^{\rm -2b}~+~ {\rm c_i}\cdot\delta F_{\rm gal}^2(\lambda).
\end{equation}
We explore the parameter space by least-squares fitting the available data from Akari and Spitzer (at $1^\prime-50^\prime$)
%\footnote{We include small scales down to 1$^\prime$ in the fitted range to appropriately account for the contribution of faint galaxies whereas the clustering signal of fluctuations is defined in the $3^\prime-30^\prime$.}
in $10^6$ Markov-Chain Monte Carlo steps. We use Metropolis-Hasting acceptance with chain burn-in of $10^5$ steps and priors on the parameters $-1<\log a<1$, $-3<\log b<3$, $-2<\log c<5$.  As each measurement has different source detection threshold we allow the last term in the above Equation to vary at each measurement depth such that $c_i$ is a free parameter in six different measurements (making the effective number of parameters eight). For the first term, we obtain $a=1.79^{+0.38}_{-0.41}$ and $b=1.77^{+0.68}_{-0.51}$. We display the 68\% confidence regions of the best-fit in Figure \ref{fig_measurements} for which we obtain ${\rm \chi^2/dof=1.62}$. The quality of the fit is limited by the slight disagreement in the data of \citet{Cooray12b} versus \citet{Kashlinsky12}, particularly at 4.5\mic\ where our single power-law spectral slope favors the latter. It is worth noting that the best-fit excess fluctuation signal, $\sim \lambda^{-1.8}$, does not exhibit a Rayleigh-Jeans type spectrum, $\sim \lambda^{-3}$ as indicated by measurements of the net large-scale signal. The reason is the contribution of remaining low-$z$ galaxies: 1) because the abundance of low-$z$ galaxies is greater at the typical detection thresholds at $<2$\mic\ i.e. their number counts are steeper, and 2) because experiments at smaller wavelengths happen to be shallower than those at $>2$\mic\ such that less of the low-$z$ galaxy contribution is removed. The SED of the best-fit excess can be seen in Figure \ref{fig_mix}.
Integrating the first term in Equation \ref{eqn_fit} over 2--5\mic\ we obtain the excess fluctuation
\begin{equation} \label{eqn_deltacib}
  \delta F_{\rm CIB}~=~\left[ \int_{\rm 2\mu m}^{\rm 5\mu m} \delta F^2 \frac{d\lambda}{\lambda} \right]^{1/2} = 0.072 ^{+0.023}_{-0.020} ~{\rm nW~m^{-2}sr^{-1}}.
\end{equation}
This compares well with the analytical estimate in \citet{Kashlinsky15}. We will refer to this value of $\delta F_{\rm CIB} =0.072$ \nW\ as our requirement for any high-$z$ model to reproduce the data in Figure \ref{fig_measurements}. We also make our equations directly scalable with $\Delta_{\rm cl}$ such that the associated CIB flux $F_{\rm CIB}=\delta F_{\rm CIB} \Delta_{\rm cl}^{-1}$ can be easily compared with measurements of the isotropic CIB, both direct and those derived from TeV blazars.

At a given near-IR wavelength, the isotropic CIB flux is related to the comoving specific emissivity $j_\nu(z)$ per unit volume of the sources
\begin{equation} \label{eqn_Fcib}
   F_{\rm CIB} \equiv \int_{2 {\rm \mu m}}^{5 {\rm \mu m}} I_\lambda d\lambda = \frac{c}{4\pi} \int \int j_\nu(z) \frac{dt}{dz}dz \frac{d\nu}{1+z}
\end{equation}
where $\nu = \nu_{\rm obs}(1+z)$ is the rest frame frequency. If the CIB was released during the first $\Delta t \simeq$ 500 Myrs, it follows from this Equation that the luminosity density $\rho_{\rm L}\equiv \nu j_\nu$ at optical wavelengths $0.2 \lsim \lambda_{\rm CIB}/(1+z) \lsim 0.5$\mic\ must reach a representative value of \citep[see also ][]{KAMM3}
\begin{equation} \label{eqn_rhoL}
  \rho_L = 1.5 \times 10^9 L_\odot ~ {\rm Mpc^{-3}}   \left( \frac{\Delta_{\rm cl}}{0.1} \right)^{-1}  \left( \frac{1+z}{10} \right) \left( \frac{\Delta t}{500~{\rm Myr}} \right)^{-1}
\end{equation}
where we have substituted $\Delta_{\rm cl}  = \delta F_{\rm CIB}/F_{\rm CIB}$. This luminosity density is 
%not only higher than the local $z=0$ value but also 
notably higher than the output during the peak of star formation history at $z\sim 1-2$. If the CIB fluctuations originate at high-$z$, the luminosity density must have been substantially higher in the early universe regardless of the nature of the sources. In the following Sections we investigate the basic astrophysical requirements for both stellar and accretion powered emission.

\section{ Stellar Sources } \label{sec:stars}

\subsection{ High-$z$ Galaxy Populations }

An increasing number of galaxies are being detected out to $z\sim 10$ as the deep Hubble program are being pushed to the limits of the instrument capabilities \citep{Bouwens07,McLure09,Ellis13,Finkelstein14,Bouwens14}. While these galaxies compose merely the tip-of-the-iceberg at these redshifts, it is of interest to estimate the expected contributions of the entire population to the unresolved CIB. Here, we present a simple forward evolution model of the conditional luminosity function which we tune to fit observations at $z\sim 8-10$. We assume that galaxies form in halos with a SFR proportional to their collapse rate at $z$
\begin{equation} \label{eqn_sfr}
  \dot{\rho}_\star(M,z) = f_\star\frac{\Omega_b}{\Omega_M}\frac{d}{dt} Mn(>M_{\rm min},t)
\end{equation}
where $f_\star$ is the average fraction of baryons in collapsed halos that are processed into stars and $n(M,t)$ is the evolving halo mass function, which we adopt from \citet{Tinker08}. In order for star formation to take place, halos must reach a sufficient size, $M_{\rm min}$, to shock heat the baryons to the virial temperate allowing for efficient gas cooling. 
 The relation between the mass of a halo and its virial temperature is 
$ M_{\rm min} = 3\times 10^6 M_\odot \left( \frac{T_{\rm vir}}{2000K}\right)^{3/2} \left(\frac{1+z}{10}\right)^{-3/2}$, 
where we have assumed a mean molecular weight of 1.22 for neutral primordial gas composed of H and He. For the discussion of high-$z$ galaxies we will adopt the limit of $T_{\rm vir}=40,000 K$, somewhat higher than the threshold for cooling via atomic hydrogen. Star formation in smaller halos will be addressed in the following subsection.
\begin{figure}
\begin{center}
      \includegraphics[width=0.48\textwidth]{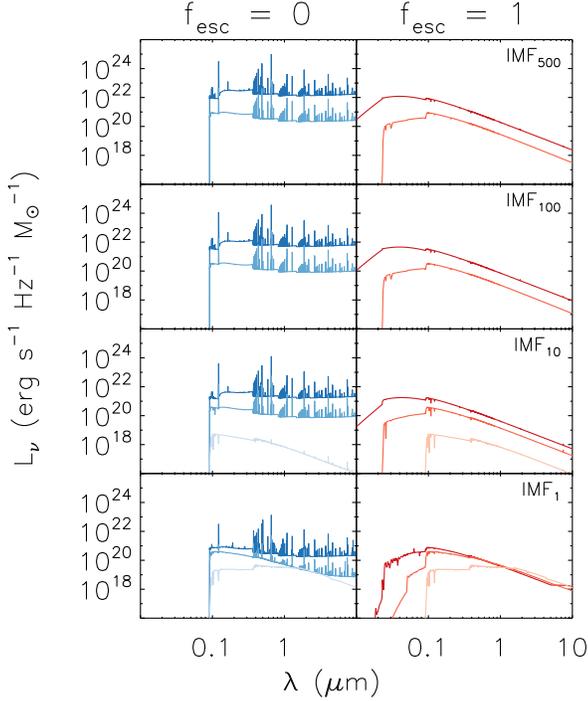}
      \caption{ The single age stellar population spectra from the {\it Yggdrasil} code. The three rows show energy spectra for different metal-free/poor IMFs. From top to bottom panels: metal-free $\sim 500 M_\odot$ stars radiating close to the Eddington luminosity, heavy IMF with characteristic mass $\sim 100 M_\odot$, intermediate IMF $\sim 10M_\odot$ and metal-poor stars with a normal Kroupa IMF (see text for details). The left and right panels show the SSP spectra for the two limiting cases of $f_{\rm esc}=$ 0 and 1 respectively. The three curves in each of the panels shows the single age population at 0 and 3.6 Myr (top two) and 0, 10, 100 Myr (bottom two). }
\label{fig_ygg}
\end{center}
\end{figure}
\begin{figure}
\begin{center}
      \includegraphics[width=0.48\textwidth]{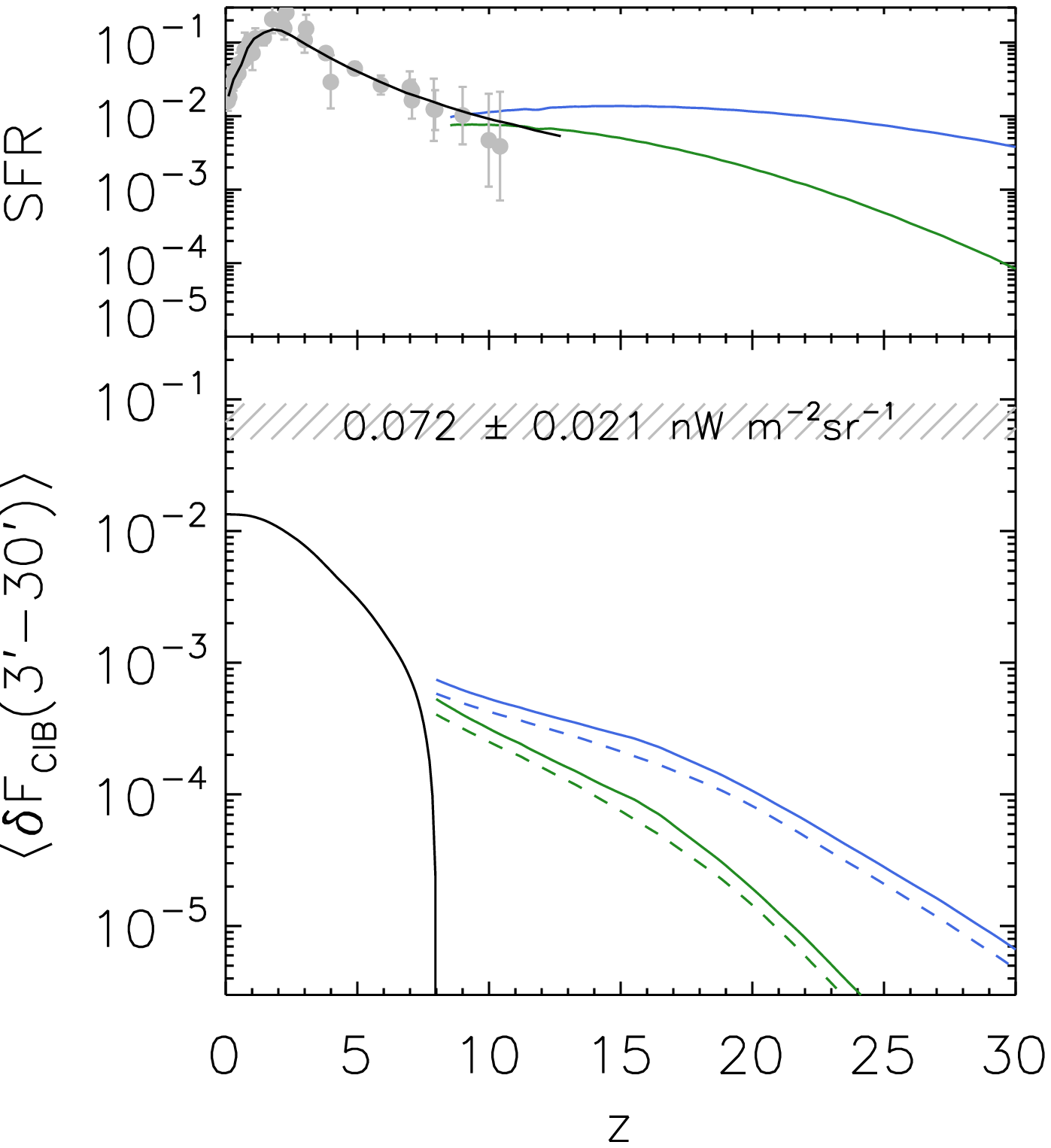}
      \caption{ {\it Upper:} The SFR in halos ($T_{\rm vir}>4\times 10^4$K, green) and minihalos ($T_{\rm vir}>10^3$K, blue). The SFR data points, compiled in \citet{Robertson15}, are derived from the UV luminosity function integrated down to $0.001L^\star$. The solid line shows the maximum likelihood SFR history from the same work. These connect well with our modeled SFR at higher-$z$ assuming $f_\star=0.005$ {\it Bottom:} The build-up of large scale averaged CIB fluctuations from galaxies forming stars with IMF$_1$ and efficiency $f_\star=0.005$. The solid and dashed lines correspond to $f_{\rm esc}=$ 0 and 1 respectively. Fluctuations from low-$z$ galaxy populations remaining after removing galaxies brighter than $m_{\rm 3.6\mu m}=25$ are shown as the black solid line \citep{Helgason12}. The reference level of the observed large scale fluctuations are shown as the hatched region (grey). }
      \label{fig_p2}
\end{center}
\end{figure}
Luminosity and spectra are assigned to star forming galaxies using the {\it Yggdrasil} model, a population synthesis code designed to model high-$z$ systems containing varying mixtures of PopII and III stars \citep{Zackrisson11}. This model includes the nebular contribution from photoionized gas and extinction due to dust, with single age stellar population (SSP) taken from \citet{Leitherer99} (PopII) and \citet{Schaerer02} (PopIII, with no mass loss). Because observed Lyman break galaxies show no evidence of anything other than metal-enriched star formation, we adopt a universal Kroupa IMF ($0.1-100M_\odot$) with a metallicity of $Z=0.0004$, characteristic of PopII stars (we call this IMF$_1$ in the discussion below and Figure \ref{fig_ygg}). The volume emissivity of halos with masses between $M$ and $M+dM$ at $z$ is obtained by convolving the SSP with the star formation occurring prior to $z$
\begin{equation}
     j_\nu(M,z) = \int_z^\infty \dot{\rho}_\star (M,z^\prime) \mathcal{L}_\nu(t_z-t_{z^\prime}) \frac{dt}{dz^\prime} dz^\prime
\end{equation}
where $\mathcal{L}_\nu(t_{\rm age})$ is the aging spectral template shown in Figure \ref{fig_ygg}. At any given epoch, each halo includes the instantaneous emission from newly formed stars as well as older populations from earlier episodes of star formation. The luminosity function is $\Phi(L)dL = \Phi(M)dM$ where the relation between mass and luminosity is $L(M) = \nu j_\nu(M,z)/(Mn(M,z))$. The CIB flux production history seen at frequency $\nu_{\rm obs}$ as a function of halo mass is 
\begin{equation}
  f(M,z) = \frac{c}{4\pi} j_\nu(M,z)\frac{\nu}{1+z}\frac{dt}{dz}
\end{equation}
where $j_\nu$ is evaluated at $\nu=\nu_{\rm obs} (1+z)$. The net CIB is then simply $F_{\rm CIB} = \int \int f(M,z) dMdz$. This can be used to derive the angular power spectrum of CIB fluctuations via projection of the source clustering \citep{Limber53}
\begin{equation} \label{eqn_limber}
  P(q) =  \int \frac{H(z)}{c d^2(z)} \left[ \int f(M,z)b(M,z) dM \right]^2 P_3(qd^{-1},z) dz
\end{equation}
where $H(z)$=$H_0\sqrt{\Omega_M(1+z)^3 + \Omega_\Lambda}$ and $d(z)$ is the comoving distance. The clustering is described both in terms of the evolving matter power spectrum in 3D, $P_3(k,z)$ \citep{EisensteinHu98} and the mass dependent halo bias, $b(M,z)$ \citep{ShethTormen01}. In this description, the bias is coupled to the brightness distribution of halos, eliminating the need for assumptions on the halo occupation number of galaxies. We also note, that variations in the cosmological parameters, in particular $\sigma_8$, affect the amplitude of the power spectrum and adds uncertainty to our modeling. Equation \ref{eqn_limber} is equivalent to the 2-halo term describing the correlation between central halos\footnote{In this paper we neglect the 1-halo term from high-$z$ sources since it is always negligible compared to the 2-halo term at the relevant angular scales ($>3^\prime$).} \citep{Seljak00,CooraySheth02}.
% Is it actually correct to show cumulative fluctuations? Because of contrast, short epoch effect...

\begin{figure*}
\begin{center}
      \includegraphics[width=0.8\textwidth]{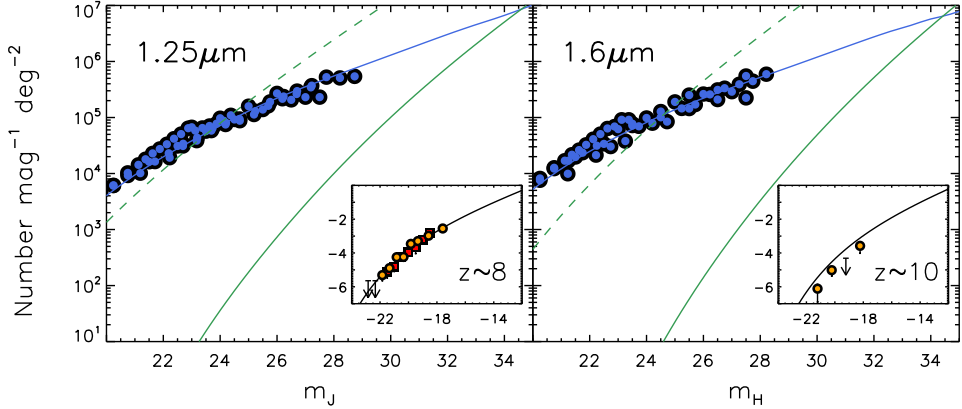}
      \caption{ Deep source counts at 1.25\mic\ (left) and 1.6\mic\ (right). The blue data points represent galaxy counts, mostly from $z\sim 1-3$ at these deep magnitudes. The galaxy counts model of \citet{Helgason12} is shown as blue solid line and extrapolated to faint magnitudes whereas the green line shows the counts for galaxies at high-$z$ ($f_\star=0.5\%$, IMF$_1$), with the Lyman dropouts placing them at $z\gsim 8$ (left) and $z\gsim 10$ (right). The comparison of low-$z$ and high-$z$ counts suggests that the faint-end of the high-$z$ galaxy population will only become significant at $\gsim 32$ mag. The dashed lines show the model that would required to reproduce the CIB fluctuations, $\delta F_{\rm CIB}$, at these redshifts. The fact that these models are in gross conflict with both the LF data and the deepest counts suggest that any such signal must either come from higher $z$ or be confined in isolated minihalos at the faint-end. The insets show the luminosity function ${\rm log( mag^{-1} Mpc^{-3})}$ vs $M_{\rm UV}$ of our default high-$z$ galaxy model ($f_\star=0.5\%$, IMF$_1$, $f_{\rm esc}=12\%$) tuned to fit measurements \citep{Finkelstein14,Bouwens14}. }
      \label{fig_lfp2}
\end{center}
\end{figure*}

The large scale CIB fluctuations resulting from our modeled high-$z$ galaxies are shown in Figure \ref{fig_p2}. We have chosen a constant star formation efficiency (in both $M$ and $z$) in such a way to obtain a good agreement with the observed star formation rate and  UV LF at high-$z$. This value we find to be $f_\star=0.005$. Figure \ref{fig_lfp2}, compares the relative abundance of high-$z$ galaxies (green) with the source density of all $z<8$ galaxies (blue). It is immediately clear that compared to current measurements, star forming galaxies with a normal IMF are underdominant at least out to 32 mag and, with these assumptions,  
%at the same time unable to 
do not reach the measured fluctuation levels (see Figure \ref{fig_p2}). This is true regardless of whether the LF is cut off at $M_{\rm UV} \simeq -14$ mag or extrapolated to much fainter systems. Simply increasing $f_\star$ further would overproduce the LF at $z \sim 8-10$ and eventually also the faint NIR number counts \citep[this was already pointed out by ][]{Salvaterra06b}. In fact, fluctuations from faint unresolved galaxies at low-$z$ produce a larger signal than high-$z$ galaxies. We therefore conclude, in agreement with \citet{Cooray12a} and \citet{Yue13a}, that the measured CIB fluctuation levels cannot be reproduced by high-$z$ galaxies with reasonable extrapolations of their evolving luminosity function.

\subsection{ Minihalos and PopIII }

There are two ways of having more light produced by stars without overproducing the observed LF at $z\sim 8-10$. First, that an epoch of more vigorous light production took place before $z\sim 8-10$; in other words, the era of CIB production already ended before this time. Second, the sources are outside the sensitivity limit of the deepest surveys i.e. they are found in numerous isolated halos that are intrinsically fainter than $M_{\rm AB}\simeq -17$ mag. This would be exhibited in a rise in the faint-end of the LF, either by greater star formation efficiency in smaller halos or by heavier IMF. However, numerical studies tend to show the opposite, that the efficiency decreases with lower masses \citep{Ricotti02a,Ricotti02b,Ricotti08,Behroozi13,OShea15}. In this subsection, we relax the assumption of a constant $f_\star =0.005$ and $T_{\rm vir}=40,000$K and explore whether greater values can be accommodated where observational constraints are not yet available. In other words, what happens if we, in addition to high-$z$ galaxies $\gsim 10^8M_\odot$, include the contribution of the first stars forming out of pristine metal-free gas in minihalos $\gsim 10^6M_\odot$?
\begin{figure}
\begin{center}
      \includegraphics[width=0.48\textwidth]{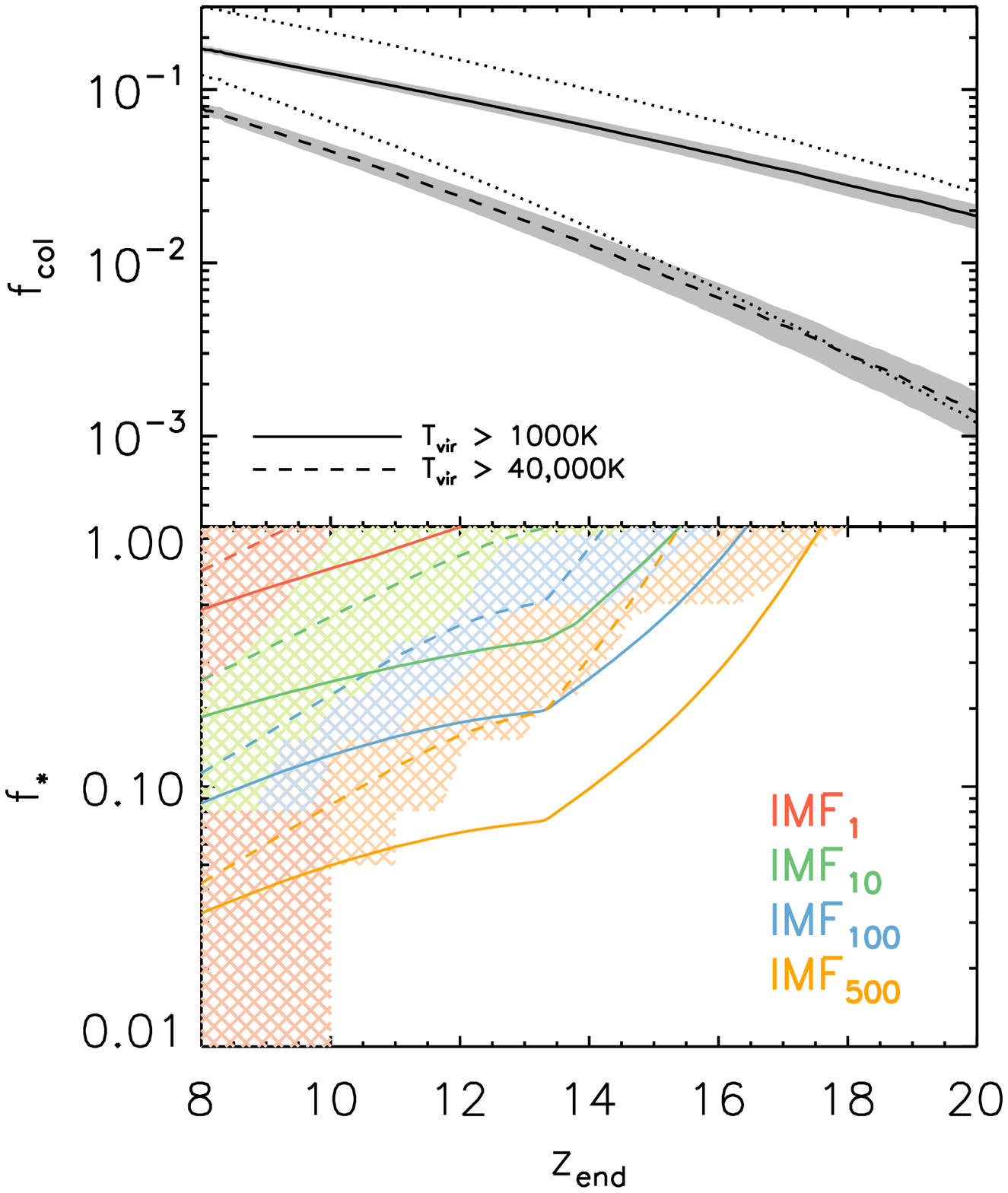}
      \caption{ 
{\it Upper}: The fraction of mass in collapsed structures as a function of redshift. Solid and dashed lines correspond to halos with $T_{\rm vir}>1,000$K ($\gsim 10^6M_\odot$) and $>40,000$K ($\gsim 10^8M_\odot$) respectively for the \citet{Tinker08} mass function. Dotted lines show the classic \citet{PressSchechter74} prediction for comparison. The grey regions show the scatter associated with the 1-$\sigma$ uncertainties in the Planck cosmological parameters. {\it Lower:} The star formation efficiency $f_\star$ required to produce the reference level of the CIB fluctuations, $\delta F_{\rm CIB}$, by a given redshift, $z_{\rm end}$. The curves assume the entire stellar population forming with IMF$_{1}$, IMF$_{10}$, IMF$_{100}$, IMF$_{500}$ (red, green,blue,orange) in all halos. The solid lines show the case where minihalos $T_{\rm vir}>1,000$K are included whereas the dashed lines include $T_{\rm vir}>40,000$K halos only. We have set $f_{\rm esc}=0$ in all cases and note energy requirements become even greater for $f_{\rm esc}>0$. The colored regions show the combination of $f_\star$ and $z_{\rm end}$ that result in the overproduction of deep NIR counts data and/or LF data for the IMFs in the same color scheme (see Figure \ref{fig_lfp2}). These regions are thus forbidden unless the bright-end of the LF is suppressed e.g. if $f_\star$ much lower in high mass halos than in low mass ones. }
\label{fig_fstar}
\end{center}
\end{figure}

The IMF of the first PopIII stars is highly uncertain but is expected to be biased towards high masses \citep{Bromm13}. It is also unclear whether stars forming under $H_2$-governed cooling (sometimes called PopIII.1) differ substantially from those forming in a second episode governed by atomic $H$ line emission (PopIII.2). We therefore consider four different IMFs of the Yggdrasil model for stars forming in halos down to the smallest minihalos equivalent to the $H_2$ cooling threshold $1,000$K
\begin{itemize}
  \item[ ] {\bf IMF$_{1}$}: standard Kroupa in the $0.1-100M_\odot$ range
  \item[ ] {\bf IMF$_{10}$}: log-normal with characteristic mass of $10M_\odot$ and dispersion of $1M_\odot$ in the $1 - 500M_\odot$ range
  \item[ ] {\bf IMF$_{100}$}:  power-law $\propto M^{-2.35}$ in the $50-500M_\odot$ range
  \item[ ] {\bf IMF$_{500}$}: all stars are 500$M_\odot$, near-Eddington
\end{itemize}
For details, we refer to \citet{Zackrisson11} and references therein. Spectral evolution templates for these IMFs are shown in Figure \ref{fig_ygg}.
%Table 1 lists the properties of the populations considered. The PopIII.1 IMF may be extreme in light of recent numerical results which indicate slightly less massive IMF of PopIII stars. 

\begin{figure*}
\begin{center}
      \includegraphics[width=0.98\textwidth]{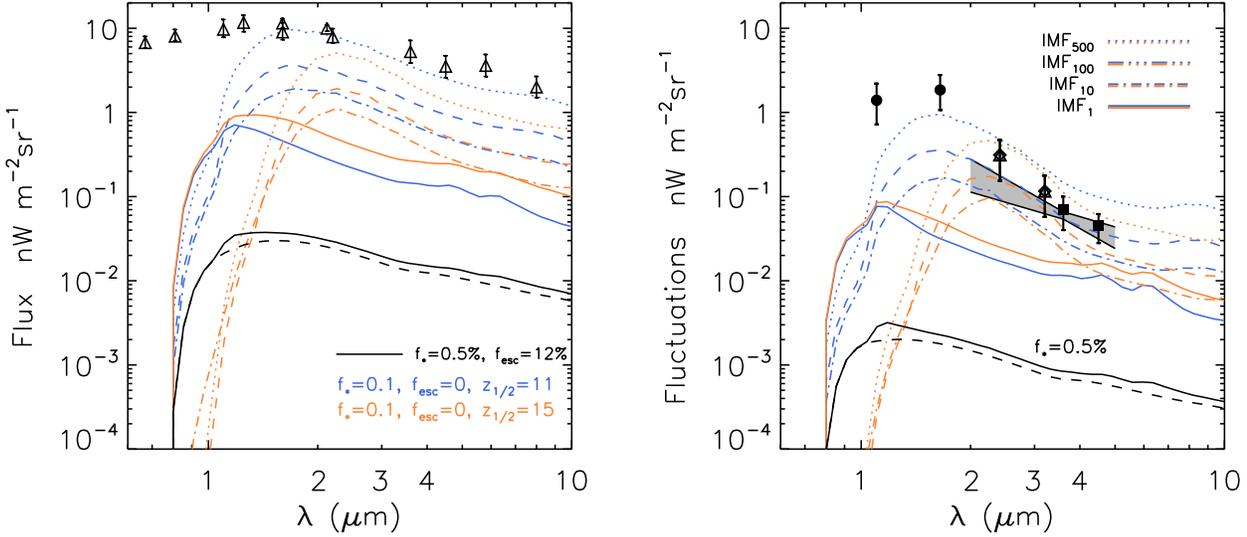}
      \caption{ The spectrum of the CIB flux, $F_{\rm CIB} = \nu I\nu$ (left) and large scale fluctuations, $\delta F_{\rm CIB} = \sqrt{q^2P/2\pi}$  (right) for models of $z>8$ early stellar emission. Fluctuations are averaged in the angular range $3^\prime - 30^\prime$. The observationally motivated model of high-$z$ galaxies, with $f_\star=0.005$ and IMF$_1$ is shown for $f_{\rm esc}=0.12$ (black solid line) and $f_{\rm esc}=1$ (black dashed) for comparison. The colored set of lines correspond to models with $f_\star= 0.1$, IMF$_{1}$ representing PopII (solid), and three cases of PopIII IMF$_{10}$,IMF$_{100}$,IMF$_{500}$ (dot-dash, dashed, dotted). The amplitude of both flux and fluctuations is directly scalable with $f_\star$. Orange corresponds to a PopIII-PopII transition with $z_{1/2}=15$ according to Equation \ref{eqn_pris}, whereas blue illustrate the case of a delayed enrichment with $z_{1/2}=11$ (see text). All star formation is ended at $z=8$. Models shown in blue and orange are for $f_{\rm esc}=0$ since $f_{\rm esc}>0.1\%$ already violates reionization constraints explained in Section \ref{sec:ion}. {\it Left:} The data points is CIB from integrated galaxy counts which is naturally much greater than the high-$z$ contribution. {\it Right:} The grey shaded region shows the best-fit CIB fluctuation excess obtained in Section \ref{sec:excess} after the contribution of low-$z$ galaxies has been accounted for. The data in the right panel is large scale averaged fluctuation measurements of \citet{Zemcov14} (filled circles), \citet{Matsumoto11} (open diamonds), \citet{Seo15} (filled diamonds) and \citet{Kashlinsky12} (filled squares). We do not show 2MASS and {\it HST}/NICMOS measurements as they do not reach sufficient angular scales for proper comparison at $3^\prime-30^\prime$. }
\label{fig_mix}
\end{center}
\end{figure*}

Figure \ref{fig_fstar} shows the average star formation efficiency required to produce the measured $\delta F_{\rm CIB}=0.072$\nW\ at any given epoch for the three IMFs considered. It is clear that under these assumptions the CIB fluctuations can only be reproduced if minihalos $\sim 10^6M_\odot$ are allowed to continuously form massive stars with a high efficiencies $f_\star>0.1$. Furthermore, deep NIR counts limit this possibility to the most heavy IMFs where much of the energy output is reprocessed into nebular emission, $f_{\rm esc}=0$ \citep[see also ][]{Fernandez10}. This is assuming that stars form continuously without mechanical or radiative feedback which is known to have strong impact on both $f_\star$ and $f_{\rm esc}$ \citep[e.g. ][]{Jeon14,Pawlik14}. In Figure \ref{fig_mix} we display the full CIB model with a simple description of chemical feedback which models the PopIII--PopII transition in terms of semi-analytic modelling of SN winds \citep{FurlanettoLoeb05} which we adopt from \citet{GreifBromm06}
\begin{equation} \label{eqn_pris}
  \dot{\rho}_{\rm PopIII} = p_{\rm pris}(z)\dot{\rho}_\star
\end{equation}
where $\dot{\rho}_\star$ is the net SFR from Equation \ref{eqn_sfr} and $p_{\rm pris}$ is the fraction of collapsed objects that are still chemically pristine. We display two cases of PopIII--PopII transition in Figure \ref{fig_mix}, standard and delayed enrichment corresponding to $z_{1/2}=11$ and $15$ respectively where we have defined $z_{1/2}$ as the redshift at which half of all star formation is in metal-enriched mode\footnote{The enrichment histories are modified by varying the radiation loss parameter $K_w^{1/3}$ \citep[see ][]{GreifBromm06}.}. The PopII assumes IMF$_1$ whereas we display the cases where PopIII takes on IMF$_{10}$, IMF$_{100}$, and IMF$_{500}$. Only the most extreme models ($f_\star>0.1$, $f_{\rm esc}\ll 1$, IMF$_{100,500}$) come close to the fluctuation excess shown in the right panel as a grey region.

\subsection{ Ionizing Photons } \label{sec:ion}

High star formation efficiencies coupled with a heavy IMF imply a vigorous production of ionizing photons ($>13.6$ eV). These photons must either be absorbed locally by neutral gas within the halo, or escape and be absorbed by still neutral parts of the IGM. The injection rate of ionizing photons can be derived self-consistently from the SFR combined with the stellar population synthesis models at $z>10$
\begin{equation}
  \dot{n}_{\rm ion}(z) = \int_{\rm >13.6eV/h} \int \frac{j_\nu(M,z)}{h\nu} dM d\nu.
\end{equation}
For the production of ionizing photons at later times $z<10$, we adopt the prescription of \citet{Robertson15} where $\dot{n}_{\rm ion}$ is derived from fitting the measured star formation rate history (see Figure \ref{fig_p2}).
We calculate the ionization fraction of the IGM, $x_{\rm ion}$, by solving
\begin{equation}
  \dot{x}_{\rm ion} = \frac{f_{\rm esc}\dot{n}_{\rm ion}(z)}{ \langle n_H \rangle } - \frac{x_{\rm ion}}{t_{\rm rec}}
\end{equation}
where $\dot{n}_{\rm ion}(z)$ is the production rate of intrinsic ionizing photons per comoving volume and $\langle n_H \rangle$ is the average comoving number density of hydrogen. The recombination timescale is $t_{\rm rec}= \left[ C \alpha_B n_H(1+Y/4X)(1+z)^3\right]^{-1}$ where $C=\langle n_H^2 \rangle /\langle n_H \rangle ^2$ is the clumping factor of ionized hydrogen, $\alpha_B(20,000{\rm K})$ is the case-$B$ recombination coefficient \citep{Hummer94} and $X=0.76$ and $Y=1-X$ are hydrogen and helium abundances. This allows us to derive the Thompson optical depth to the electron scattering
\begin{equation}
\tau_e =  c \sigma_T \langle n_H \rangle \int x_{\rm ion}(z) \left(1+ \frac{\eta Y}{4X} \right)(1+z)^3 \frac{dt}{dz}dz
\end{equation}
where we assume Helium is singly ionized at $z>3$ ($\eta=1$) and fully ionized at later times ($\eta=2$).  This can be compared with the latest values inferred from CMB polarization measurements from Planck $\tau_e=0.066\pm 0.012$ \citep{Planck15}, which were found to be somewhat lower than earlier results from WMAP $\tau_e=0.088\pm 0.014$ \citep{Hinshaw13}.
\begin{figure*}
  \hfill
  \begin{minipage}{.49\textwidth}
    \begin{center}
      \includegraphics[width=1.\textwidth]{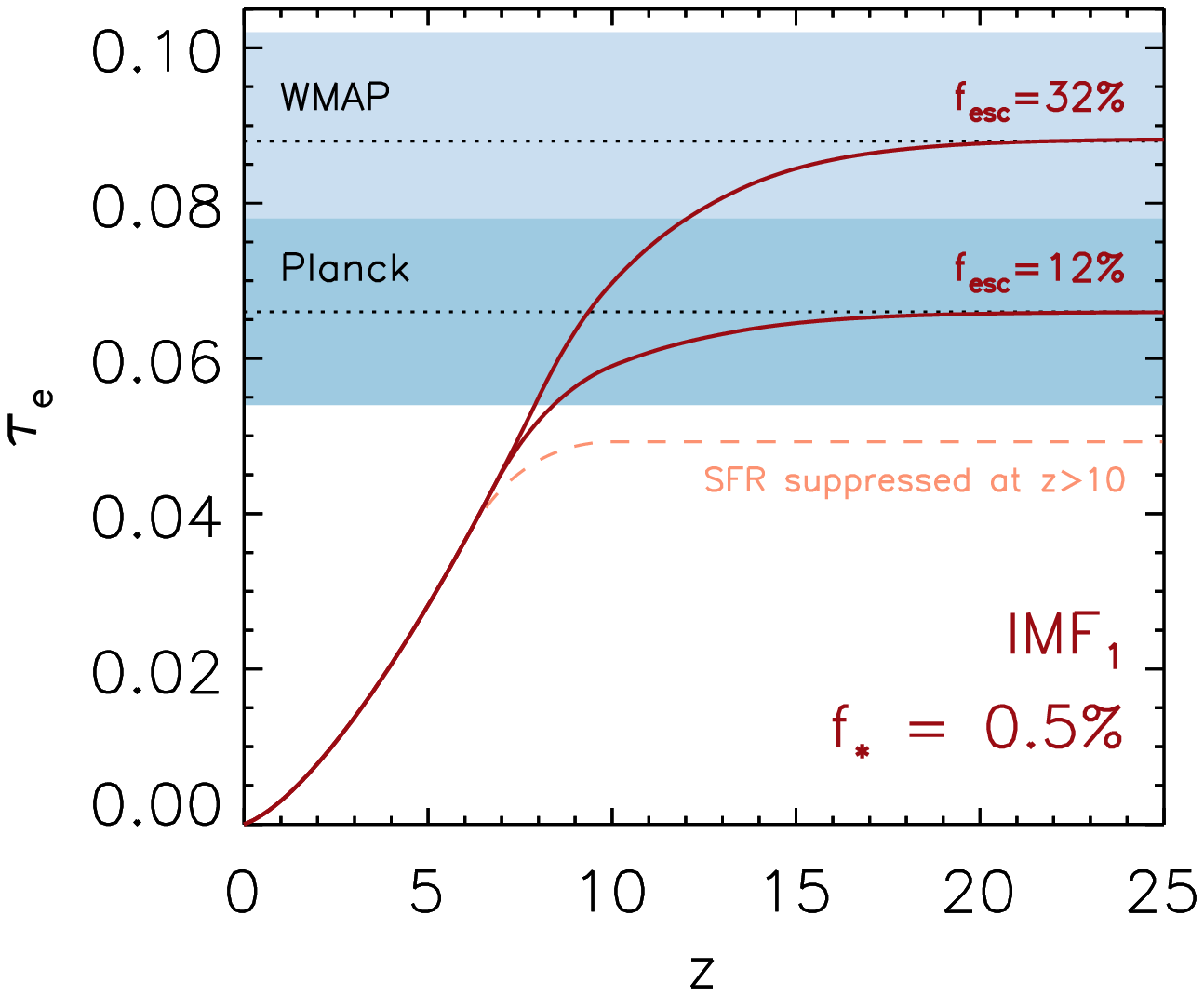}
    \end{center}
  \end{minipage}
  \hfill
  \begin{minipage}{.49\textwidth}
    \begin{center}
      \includegraphics[width=1.\textwidth]{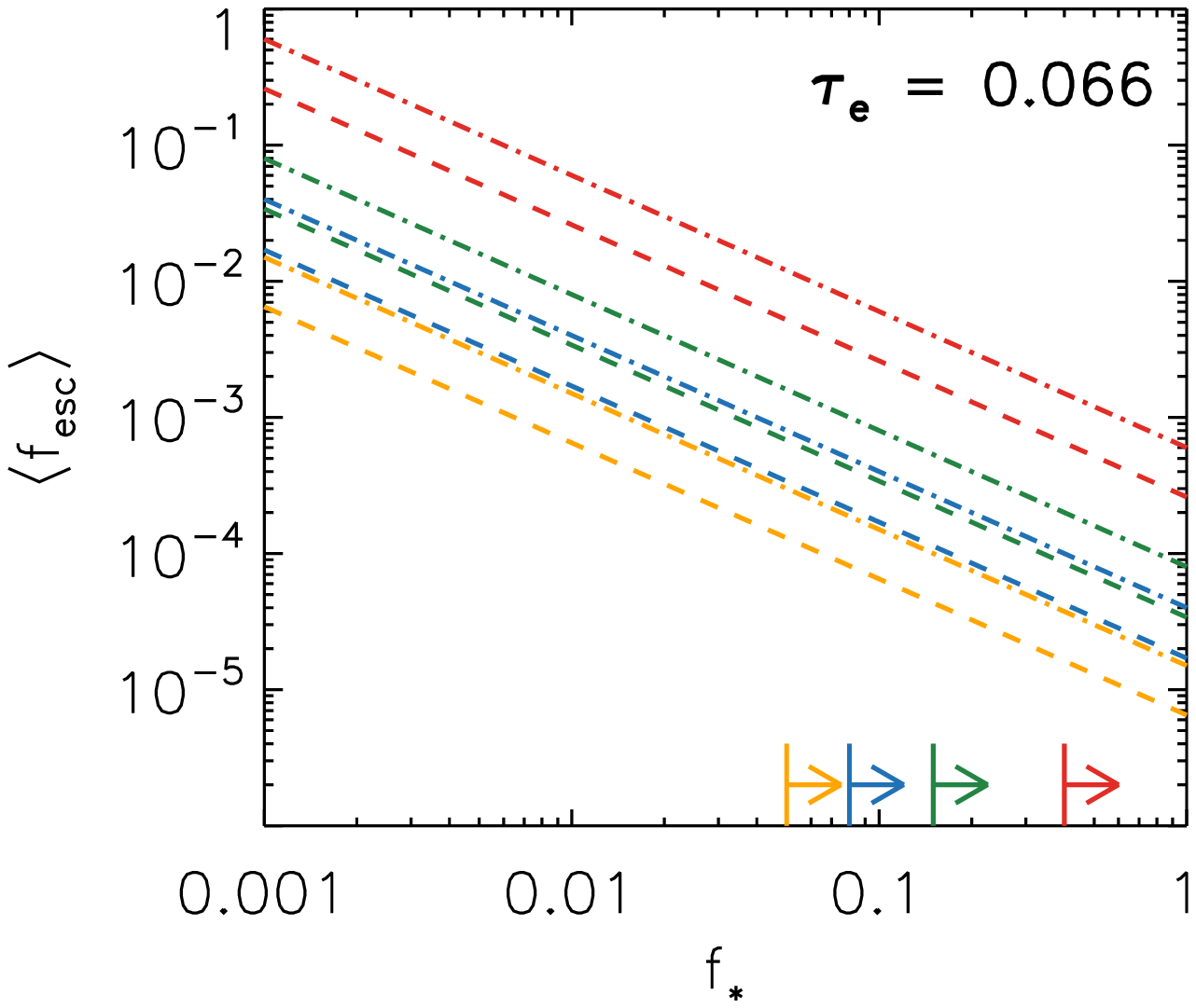}
    \end{center}
  \end{minipage}

      \caption{ {\it Left:} The Thomson electron scattering optical depth integrated over redshift for our standard model of galaxy formation at $z>10$. The $z<10$ contribution shown in light red is derived from the measured star formation rate history shown in Figure \ref{fig_p2} \citep[see ][]{Robertson15}. The red lines show the escape fraction tuned to match $\tau_e$ from Planck (lower) and WMAP (upper) which are indicated as horizontal dotted lines with blue regions showing the 1$\sigma$ uncertainty. {\it Right:} The mean escape fraction versus star formation efficiency in all collapsed halos for a fixed value of $\tau_e=0.066$ assuming the four IMFs in the same color scheme as in Figure \ref{fig_fstar}. Dashed corresponds to a cooling threshold of $T_{\rm vir}>1,000$K ($\sim 10^6M_\odot$) and dot-dashed to $T_{\rm vir}>40,000$K ($\sim 10^8M_\odot$). In order to reproduce $\delta F_{\rm CIB}$ and at the same time be consistent with reionization, the combination of ($f_{\rm esc}$,$f_\star$) must be concentrated in the lower right corner. The arrows denote the minimum required $f_\star$ are taken from Figure \ref{fig_fstar} (lower) at $z=10$. }
    \label{fig_reion}
  \hfill
\end{figure*}
%\begin{figure*}
%\begin{center}
 %     \includegraphics[width=0.8\textwidth]{fig_fesc.eps}
  %    \caption{ The mean escape fraction versus star formation efficiency in all collapsed halos ($>10^3K$) for a fixed electron scattering optical depth, $\tau_e=0.05$ reached at $z_{\rm end}=10$ (solid) and $z_{\rm end}=14$ (dot-dashed). The three panels correspond to different clumping factors, $C=3,5,10$ (left, center, right). In order to reproduce $\delta F_{\rm CIB}$ and at the same time be consistent with the reionization, $f_{\rm esc}$-$f_\star$ parameters must be below the lines, outside the colored regions for each IMF (same color scheme as in Figure \ref{fig_fstar}). The arrows denote the minimum required $f_\star$ are taken from Figure \ref{fig_fstar} (lower) at $z=10$. This shows that ionizing photons must be nearly completely absorbed. }
%\label{fig_fesc}
%\end{center}
%\end{figure*}
In Figure \ref{fig_reion} (left), we show the integrated optical depth to electron scattering, $\tau_e$, as a function of redshift for our preferred IMF$_1$ model of high-$z$ galaxies. Forcing the model to reproduce the latest Planck measurements requires $f_{\rm esc}=12\%$ and $5\%$ for minimum host halo mass corresponding to $40,000$K ($\sim 10^8M_\odot$) and $10^3$K ($\sim 10^6M_\odot$) respectively. For the somewhat higher WMAP-derived $\tau_e$ these values are $f_{\rm esc}=32\%$ and $12\%$ respectively. We caution that, since we did not include any feedback effects which tend to surpress the star formation efficiency at small host halo masses, the average escape fraction may be somewhat higher.

For a given SFR and IMF, the injection rate of ionizing photons into the IGM is simply proportional to $f_\star$ and inversely proportional to $f_{\rm esc}$. Figure \ref{fig_reion} (right) shows the combination of $f_\star$ and $f_{\rm esc}$ that give a fixed $\tau_e=0.066$ for our four IMFs. The Figure shows that in order to maintain the required $f_\star$ and at the same time avoid reionizing the universe too early (and overproducing $\tau_e$), one must maintain a very low average escape fraction $f_{\rm esc}\lsim 0.1\%$ at $z>10$. It follows that the $\gsim 99\%$ of the ionizing photons need to be absorbed by the local gas which itself can only constitute $(1-f_\star)<90\%$ of the available baryons in these small halos. Several studies modeling the propagation of ionization fronts within halos hosting the first galaxies show that the escape fraction increases towards smaller mass \citep{RicottiShull00,Johnson09}. For star formation efficiencies $f_\star>10^{-3}$, \citet{FerraraLoeb13} find that ionizing photons escape very easily with $f_{\rm esc}\simeq 1$ across a wide range of halo mass. Only the most massive halos, $\gsim 10^8 M_\odot$ are able to confine their UV photons effectively. The escape fraction is also expected to increase towards higher redshifts making it hard to justify large $f_\star$ in high-$z$ minihalos \citep{Kuhlen12,Mitra13}. There are recent indications from determinations of the high-$z$ LF that the star formation efficiency is actually increasing at early times \citep{Finkelstein15}. While it is not clear whether this trend extends to the lowest mass halos where cooling relies on $H_2$, observations of the stellar mass function in local dwarf systems suggest that small atomic cooling halos at high-$z$ could indeed be very efficient sites of star formation \citep{Madau14}.

\subsection{ Supernova contribution }

The contribution of supernova to the CIB was briefly discussed by \citet{CoorayYoshida04} who argued that SNe would remain subdominant with respect to the stellar contribution. The net energy radiated in a core-collapse SN is $E_{\rm SN} \sim 10^{51} {\rm erg}$ out of which only 0.1-1\% emerges as electromagnetic radiation. For comparison, a star burning 10\% of its initial hydrogen emits $E_\star = 10^{52} (M_\star/M_\odot) {\rm erg}$ over its lifetime, out of which only a fraction of this energy will end up in the CIB, depending on the spectrum and redshift. A crude calculation can be made based on our model by assuming that every $>8M_\odot$ star explodes as a supernova for our three IMFs above. Even if every SNe inputs a generous $10^{49}$ergs into the CIB, the contribution is always $<0.1\%$ of the stellar main-sequence contribution. It is therefore safe to ignore SNe as a major CIB contributor even in the case of some unusually bright SN-types (PISN, hypernova) that have been proposed \citep{Barkat67,Iwamoto98}.

\subsection{ Fossils of the first galaxies }

Around the time of reionization, it is expected only systems with sufficient mass ($\gsim 10^8-10^9M_\odot$) will be shielded from photoheating and continue to form stars. The majority of the already collapsed low-mass systems can nevertheless continue to contribute to the CIB through ageing stellar populations that continue to radiate throughout cosmic time. Locally, these systems are referred to as the fossils of the first galaxies \citep{RicottiGnedin05,BovillRicotti09}. These will not be as intrinsically bright as their high-$z$ progenitors but can in principle accumulate substantial CIB by emitting at much lower-$z$ and for much longer span of time. Our formalism outlined above accounts for the ageing of populations and allows us to estimate the fossil contribution at later times. Approximating reionization as a step function, we turn off the SFR at $z=10$ and allow the galaxies to passively evolve to $z=0$. Not all fossil stars will contribute to the source-subtracted CIB as many systems will be incorporated into larger halos hosting star forming galaxies that are masked and subtracted in a fluctuation study \citep{Bovill11a,Bovill11b}. For simplicity we do not account for any subtraction of the fossil contribution and the calculation presented here includes all star-forming progenitors at $z>10$. This should therefore be taken as an upper limit to the source-subtracted CIB for pre-reionization fossils. Note, however that this simple picture has recently been challenged by recent evidence showing that many dwarf systems continued to form stars even well after the universe was reionized \citep[e.g. ][]{Weisz14}.
\begin{figure}
\begin{center}
      \includegraphics[width=0.5\textwidth]{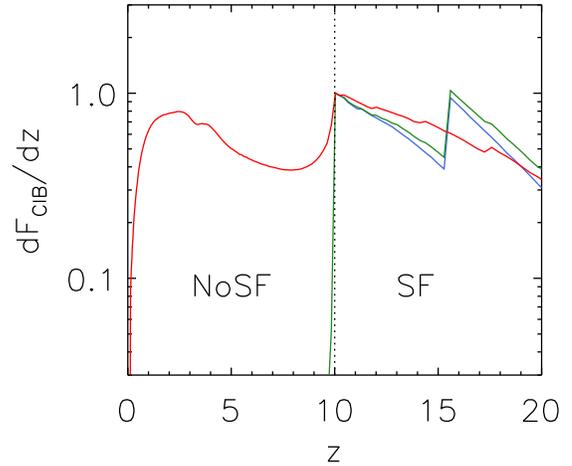}
      \caption{ The integrated 2--5\mic\ flux production when star formation is turned off at $z=10$, normalized to unity at $z=10$ (arbitrary units). The red curve corresponds to a standard IMF$_1$ containing ageing low-mass stars in all $T_{\rm vir}>10^3$K ($\gsim 10^6M_\odot$) halos. The green and blue lines show the heavier IMF$_{10}$ and IMF$_{100}$ respectively where the CIB contribution drops quickly after star formation is shut off as heavier stars die. The curves have been normalized at $z=10$ for easier comparison. }
\label{fig_fossil}
\end{center}
\end{figure}

Figure \ref{fig_fossil} shows the emission history of star forming galaxies ($z>10$) and their remnant fossils ($z<10$). Shortly after SFR is set to zero, the flux from IMF$_{10}$ and IMF$_{100}$ drops quickly as more massive stars die out but for IMF$_{1}$ the contribution of the remaining low-mass stars  gradually rises towards lower-$z$ due to the $(1+z)^{-1}$ factor in Equation \ref{eqn_Fcib}. Interestingly, when integrated over redshift the CIB contribution from fossils is $\sim$70\% of the net contribution whereas high-$z$ galaxies make up only $\sim 30$\%. This suggests that the fossil contribution should be included in the net energetics of background radiation. However, considering that the low-$z$ part is an {\it upper} limit for the fossil contribution, this is still insufficient to account for the measured $\delta F_{\rm CIB}$. It will only become important if the corresponding high-$z$ contribution is shown to be significant.

%It is also worth noting, that although CDM predicts an abundant widespread population of dwarfs, they are preferentially reside in denser environments such as in galaxy groups and as satellites around host galaxies. However, the fact that the CIB clustering component does not seem to spatially correlate with masked galaxies and their extended wings already argues against a substantial contribution from undetected dwarfs.

\section{ Accretion Powered Emission } \label{sec:bhs}

\begin{figure}
\begin{center}
      \includegraphics[width=0.48\textwidth]{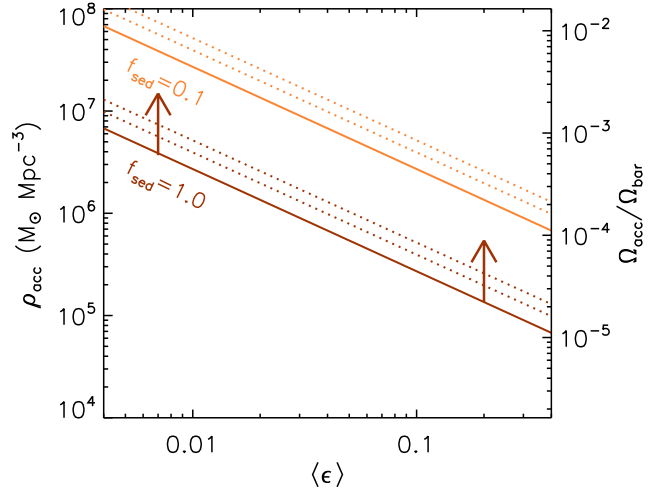}
      \caption{ The requirements for accreted mass density and time averaged efficiency to produce the reference $\delta F_{\rm CIB}$ level. The lower limit corresponds to $f_{\rm sed}=1$ when all the energy emitted ends up being observed in the NIR. The requirements increase quickly as the spectrum becomes broader $f_{\rm sed}<1$. The solid lines correspond to $z_{\rm end}=10$ whereas the dotted lines show $z_{\rm end}=15,20$ (lower/upper). The right vertical axis shows the accreted density as a fraction of the baryonic density.   }
      \label{fig_acceff}
\end{center}
\end{figure}

CIB originating from accretion onto black holes offers a more energetically favorable scenario due to the more efficient conversion of mass to energy. Such a scenario is particularly attractive for explaining the CIB in light of the observed coherence with the CXB \citep{Cappelluti13} and the expectation that the progenitors of supermassive black holes (SMBH) in the local universe were growing rapidly during this epoch \citep[see ][ for review]{VolonteriBellovary12}. With an average radiative efficiency of accretion $\epsilon = L/\dot{M}c^2$, a black hole will increase its mass by $\dot{M} = (1-\epsilon(M,t))\dot{M}_{\rm in}$ where $\dot{M}_{\rm in}$ is the net incoming mass accretion rate. Throughout this section we will assume a constant efficiency, $\epsilon(M,t)=$const with a caveat that relaxing this assumption may alter the numerology that follows. The net energy density radiated by a black hole population can be related to mass density accreted before redshift $z_{\rm end}$ \citep{Soltan82}
\begin{equation}
  \rho_{\rm acc}(z_{\rm end}) = \frac{1-\epsilon}{\epsilon c^2}\int_{z_{\rm end}}^\infty \frac{dt}{dz}dz \int \nu j_\nu(z) d\nu
\end{equation}
which is independent of the individual masses of the black hole population. The energy output is related to the accretion reate via $\rho_L=\epsilon \dot{\rho}_{\rm acc}c^2$ for which the CIB can be written (see Equation \ref{eqn_Fcib})
\begin{equation} \label{eqn_Fsoltan}
  F_{\rm CIB} = \frac{ c^3}{4\pi} \int_{z_{\rm end}}^\infty \epsilon \dot{\rho}_{\rm acc}dt \int_{\nu_0}^{\nu_1} \frac{  b_\nu d\nu}{1+z}
\end{equation}
 where $b_\nu$ is the average emerging spectrum, normalized such that $\int b_\nu d\nu=1$ and taken to be independent of mass and time. When the last term is roughly constant or slowly varying at early times (in particular for a blue spectrum sloping up with shorter wavelengths) the net CIB from accretion processes is simply proportional to the net mass accreted by redshift $z_{\rm end}$ i.e. $F_{\rm CIB} \propto \rho_{\rm acc}$. In other words, {\it  with these assumptions the CIB is mostly independent of both the growth history and the individual masses of the BHs}. It follows from Equations \ref{eqn_fracfluc} and \ref{eqn_Fsoltan} that the accreted mass density required to produce the CIB fluctuations by redshift $z_{\rm end}$ is roughly
\begin{equation} \label{soltan}
  \begin{split}
  \rho_{\rm acc} \simeq & 2\times 10^5  M_\odot~{\rm Mpc^{-3}} \left( \frac{\delta F_{\rm CIB}}{0.072~{\rm nW~m^{-2}sr^{-1}}} \right) \\
  &\times\left( \frac{\Delta_{\rm cl}}{0.2} \right)^{-1}\left( \frac{ \epsilon }{0.1} \right)^{-1} \left( \frac{1+z}{10} \right) \left( \frac{f_{\rm sed}}{1.0} \right)^{-1}
  \end{split}
\end{equation}
where $\epsilon$ is the time-averaged efficiency, $f_{\rm sed}$ is a bolometric correction factor i.e. the fraction of the total energy emitted that ends up being observed in the NIR (2--5\mic). This requirement is illustrated in Figure \ref{fig_acceff}. With maximal efficiency ($\epsilon=0.4$) we require a minimum of $\rho_{\rm acc} > 4\times 10^4 M_\odot {\rm ~Mpc^{-3}}$ for producing the CIB fluctuations. If this is to be reached at high-$z$, any accretion activity by these black holes at later times would inevitably grow the net mass density locked in BHs further. Given the brief cosmic time elapsed at high-$z$, this lower limit for the accreted mass is quite large in comparison with the mass density in SMBHs in the local universe $z=0$, estimated at $(4-5)\times 10^5 M_\odot {\rm Mpc^{-3}}$ \citep{Shankar04,Marconi04,Vika09}\footnote{Recent evidence suggests this could be as much as five times higher than previously estimated \citep{Comastri15}}. Other studies place additional limits on the mass density at $z\sim 6$ to $\lsim 10^4 M_\odot{\rm Mpc^{-3}}$ based on the unresolved CXB \citep{Salvaterra12}; and $\lsim 10^3 M_\odot{\rm Mpc^{-3}}$ based on the integrated X-ray emission of high-$z$ sources \citep{Treister13}. Both these limits can however be avoided if the emerging SED is significantly altered, or if BH cores are heavily obscured or the accretion is radiatively inefficient. Relaxing the assumptions of constant $\epsilon$ would likewise alter these numbers. Nevertheless, for the CIB to be entirely from high-$z$ we require an abundant radiatively efficient population of black holes established very early on, with significantly slowed/inefficient growth over the majority of the remaining cosmic time \citep[see e.g. ][]{Tanaka12}. Whether the required $\rho_{\rm BH}$ is in fact realistically attainable is a matter of debate and outside the scope of this paper. In what follows we do not need to make assumptions on the BH seed masses and growth mechanism but will instead assume that the required $\rho_{\rm acc}$ can be reached by $z=10$ and explore the observational consequences.

\subsection{ Correlation with the Unresolved Cosmic X-ray Background }

\begin{figure}
\begin{center}
      \includegraphics[width=0.48\textwidth]{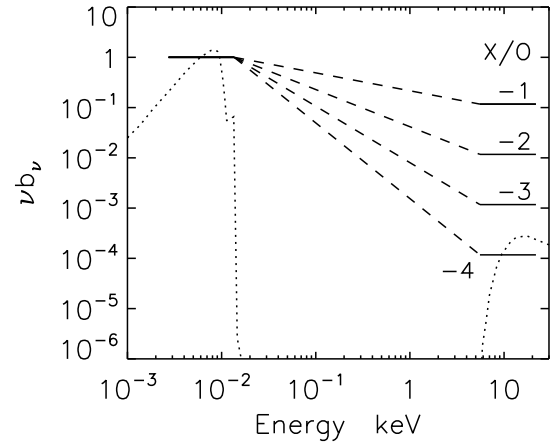}
      \caption{ A simplistic broad-band SED in the rest-frame with two contributions 1) optical/UV which is normalized to unity, and 2) a hard X-ray component which is a factor of $10^{-1}, 10^{-2}, 10^{-3}, 10^{-4}$ less energetic. For reference, the dotted curves show the emerging spectrum from the DCBH model of \citet{Yue13b} also normalized to unity in the UV/optical. These spectra are used as templates in Figure \ref{fig_bh}. }
      \label{fig_bhsed}
\end{center}
\end{figure}
\begin{figure*}
\begin{center}
      \includegraphics[width=0.95\textwidth]{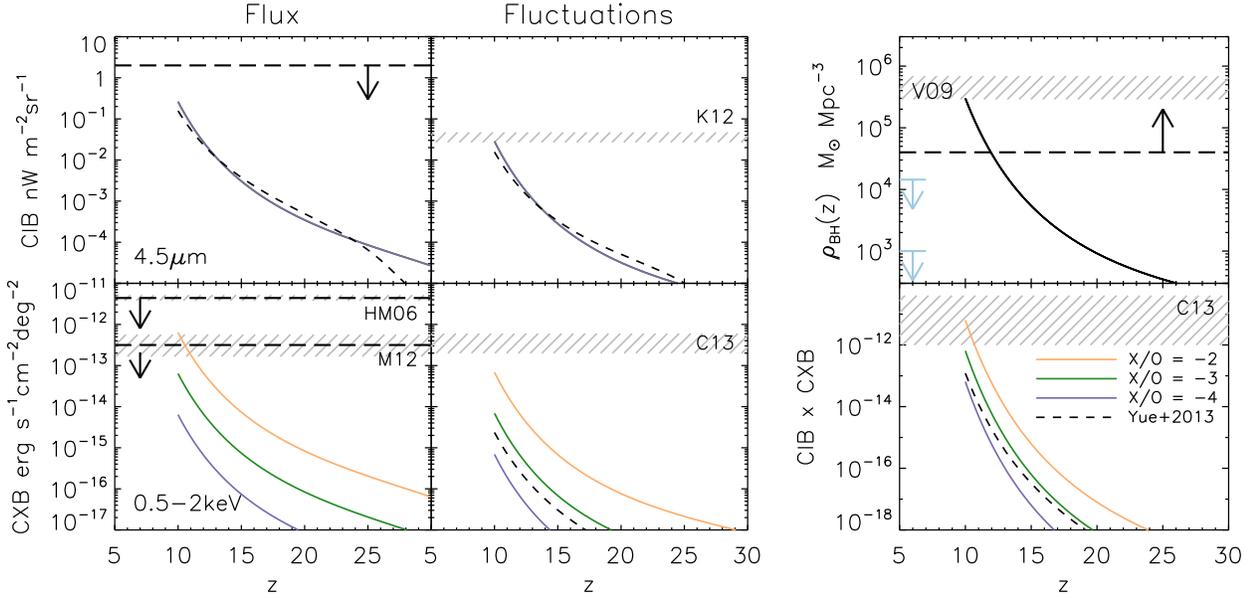}
      \caption{ {\it Left:} The cumulative CIB (upper) and CXB (lower) from black holes at high-$z$ growing according to the BH growth in the upper right panel. The $\rho_{\rm acc}$ at $z=10$ has been chosen such that the measured $\delta F_{\rm 4.5\mu m}=0.037\pm 0.010$ is reproduced \citep[marked as K12, ][]{Kashlinsky12}. The results here are mostly independent of the growth history and the individual BH masses. The upper left panel shows the resulting $F_{\rm 4.5\mu m}$ CIB where TeV upper limits from \citet{Meyer12} are also shown. The lower panels show the CXB assuming three SED templates for the $X/O$=-2,-3,-4 (yellow, green, blue) and that from \citet{Yue13b} (dashed curve) shown in Figure \ref{fig_bhsed}. Different upper limits for the CXB from \citet{Hickox06} (HM06) and \citet{Moretti12} (M13, converted to 0.5--2 keV flux) are shown as dashed lines with 1$\sigma$ uncertainties as grey regions. The levels of the large scale CXB fluctuations are shown in the lower right panel with the measured levels of \citet{Cappelluti13} (C13) shown for reference. 
{\it Right:} Accreted mass density in black holes growing at Eddington with $\epsilon=0.1$ reaching  $\rho_{\rm BH}=3 \times 10^5M_\odot{\rm Mpc^{-3}}$ by $z=10$ (solid). The dashed line shows the lower limit for $\delta F_{\rm CIB}$ to be produced according Equation \ref{soltan} in the extreme limit $f_{\rm sed}=1$ and $\epsilon =0.4$. The mass density locked in SMBHs at $z=0$ is also shown for reference as grey regions \citep{Vika09} (V09) and upper limits at $z\sim 6$ from \citet{Salvaterra12} and \citet{Treister13} in blue (upper, lower respectively). The lower panel shows the CIB$\times$CXB fluctuations in units of ${\rm erg~s^{-1}cm^{-2}nW~m^{-2}sr^{-2}}$. Notice how the $X/O$=-2 case (yellow) is the only one to reach the measured levels of \citet{Cappelluti13} and \citet{Yue13b} (dashed) appears notably below this level. }
\label{fig_bh}
\end{center}
\end{figure*}
The recently detected cross-correlation signal between the unresolved CIB (3.6 and 4.5\mic) and CXB (0.5--2keV) fluctuations was suggested to originate from black holes in the early universe \citep{Cappelluti13}. After accounting for unresolved AGN and X-ray binaries within galaxies, \citet{Helgason14} find that a tentative CIB$\times$CXB signal (2.5$\sigma$) remains at large-scales exhibiting the same shape as the CIB fluctuation power spectrum. From the definition of the spatial coherence $\mathcal{C}(q) = P^2_{\rm CIB\times CXB}/(P_{\rm CIB}P_{\rm CXB})$ and that $\langle \delta F^2\rangle \simeq q^2P(q)/2\pi$ one can write
\begin{equation} \label{eqn_coherence}
  \delta F_{\rm CIB\times CXB}^2 = \sqrt{\mathcal{C}} ~ \delta F_{\rm CIB} ~ \delta F_{\rm CXB}
\end{equation}
where $\delta F_{\rm CIB\times CXB}$ is the CIB$\times$CXB cross rms-fluctuation. Taken at face value, the large scale average (3$^\prime$-30$^\prime$) of this measurement gives $\delta F_{\rm CIB\times CXB} \sim 10^{-11}{\rm erg~s^{-1}cm^{-2}nW~m^{-2}sr^{-2}}$. With the CIB fluctuations being roughly $\delta F_{\rm 4.5\mu m} \simeq 0.04 {\rm nW~m^{-2}sr^{-1}}$ (see Section \ref{sec:excess}) we must have $\sqrt{\mathcal{C}}\delta F_{\rm CXB} \simeq 10^{-8} {\rm erg~s^{-1}cm^{-2}sr^{-1}}$ to satisfy the CIB$\times$CXB correlation. From Equations \ref{eqn_coherence} and \ref{eqn_fracfluc}, the implied CXB flux is
\begin{equation} \label{eqn_fx}
 F_{\rm CXB} =   7 \times 10^{-13} ~ {\rm erg~s^{-1}cm^{-2}deg^{-2}} \left( \frac{1.0}{\sqrt{\mathcal{C}}} \right)\left(\frac{0.1}{\Delta_{\rm cl}} \right).
\end{equation}
Since $F_{\rm CXB}$ is minimized for perfectly coherent sources, $\mathcal{C}=1$, this value represents the amount of X-ray flux associated with the CIB sources whereas it is a lower limit for the total unresolved CXB. However, this flux can be accommodated by independent measurements of the unresolved CXB in \citep{Hickox06} but is only marginally consistent with \citet{Moretti12} who find only $\sim 3 \times 10^{-13} ~ {\rm erg~s^{-1}cm^{-2}deg^{-2}}$ to be still unresolved\footnote{ We have converted the 1.5keV flux of \citet{Moretti12} to 0.5--2keV using their spectral index $\Gamma = 0.1$ }. However, \citet{Moretti12} do not include energies $<1.5$ keV in their fit for the CXB spectrum which could be the source of their lower derived 0.5--2 keV flux. On the other hand, the required $F_X$ could itself be reduced if i) the large scale clustering of the underlying sources is stronger, say with $\Delta_{\rm cl}\gsim 0.2$, or ii) if the CIB$\times$CXB level at large scales is actually lower in reality, which is possible within the uncertainties, bringing it into agreement with low-$z$ AGN and XRBs \citep{Helgason14}.

To illustrate these requirements, we display four simplistic SEDs in Figure \ref{fig_bhsed} where we assume an optical/UV component and an X-ray component that is a factor of $10^{-1}, 10^{-2}, 10^{-3}, 10^{-4}$ less energetic. We refer to these in terms of the X-ray to optical ratio, $X/O=-1,-2,-3$ and $-4$ respectively where $X/O=\log_{10} L_X/L_O$ with $L_X$ and $L_O$ integrated in the ranges $(1+z)[0.5,2]$keV and $[4.5/(1+z),0.091]$\mic\ respectively, with $z=10$ . These ratios are already significantly lower than locally observed AGN which exhibit $X/O\sim 0$ on average but a suppressed X-ray component is possible for highly obscured BHs. The ionizing photons are then absorbed by the thick gas and reprocessed into nebular emission in the UV/optical (Ly-$\alpha$, free-free), enhancing the CIB contribution \citep{Yue13b}. In this way, a larger fraction $f_{\rm sed}$ of the bolometric luminosity ends up in the CIB, requiring less accretion onto BHs (see Fig. \ref{fig_acceff}). We display the emerging SED from the highly obscured DCBH model of \citet{Yue13b} in Figure \ref{fig_bhsed} for comparison. The energy at longer IR/radio wavelengths, and shorter $\gamma$-ray wavelengths are assumed to be negligible in comparison.

Figure \ref{fig_bh} shows the cumulative CIB and CXB flux including fluctuations and CIB$\times$CXB cross fluctuations resulting from these broad-band SEDs. Because the shape of the growth history does not matter much, we simply assume a scenario in which all black holes grow at the Eddington rate with $\epsilon=0.1$, reaching a fixed accreted mass density at $z_{\rm end}=10$ ($\equiv \rho_{10}$)
\begin{equation}
  \rho_{\rm acc}(z) ~=~ \rho_{10}~{\rm exp}\left[ - \left( \frac{1-\epsilon}{\epsilon}\right) \frac{t_{10}-t_z}{t_{\rm Edd}} \right]
\end{equation}
where $t_{\rm Edd}\simeq 0.45$ Gyr is the Eddington time scale. The results are also not sensitive to the choice of $z_{\rm end}$ as long as we keep $\epsilon$ and $\rho_{10}$ fixed. We have chosen $\rho_{10}=3\times 10^5 M_\odot~{\rm Mpc^{-3}}$ such that the $\delta F_{\rm CIB}$ is reproduced. We find that the indicative large-scale CIB$\times$CXB signal can only be attained when $X/O\sim -2$. At the same time, the unresolved CXB implied by $X/O\gsim -2$ is in slight tension (1-$\sigma$) with the most stringent limits measured by \citet{Moretti12} (see Fig. \ref{fig_bh} bottom left). The spectrum of \citet{Yue13b} having $-4<X/O<-3$ is shown as a dashed line in Figure \ref{fig_bh}, and appears notably below the CIB$\times$CXB requirement. This means that the \citet{Yue13b} model actually falls short of the X-ray energy implied by the coherent CXB levels but is nevertheless consistent with the CIB levels.

It is important to emphasize that the sources producing the CIB and CXB are not necessarily the same physical emitters and are in general a mixture of stars and BHs sharing the same large scale structures. However, if they are predominantly BHs emitting in both optical and X-rays they will be accompanied by ionizing UV emission. For AGN spectra, this contribution is estimated by connecting the optical and X-ray part of the spectrum with a power law slope, shown as dashed lines in Figure \ref{fig_bhsed}. If we adopt the measured amplitudes of the CIB and CXB fluctuations we obtain a slope ranging from $d\log{(F_O/F_X)}d\log{E}=-0.5$ to $-0.7$ and the resulting ionization rate exceeds the recombination rate in all cases such that $t_{\rm reion} \ll t_{\rm recomb}$. This means that in the presence of such production rate of ionizing photons at $z\gsim 10$, the Universe will reionize in the matter of $\sim$tens of Myr. For the CIB$\times$CXB signal to arise from the same type of sources therefore requires the suppression of UV photons escaping into the IGM ($f_{\rm esc} <1$) e.g. the black holes to be ultra-obscured \citep{Ricotti05a,Comastri15}. This would be easier with unconventional BHs such as DCBHs rather than X-ray binaries from stellar remnants which are much less obscured and are limited in growth by the mass and lifetime of the companion star \citep[e.g. ][]{Yue14,Tanaka14,Pacucci15}. Also, for an abundant population of small BHs, a proportionally greater fraction of the net mass goes into the seed of the gravitational collapse and does not contribute to $\rho_{\rm acc}$.

\section{ Summary and discussion }

In this paper, we assessed the possibility that the established clustering signal in the source-subtracted CIB fluctuations at near-IR originates in the early universe. We consider emission from star formation and black hole accretion during the first 500 Myrs and establish the physical requirements and consequences to reach the measured CIB values. Our results can be broadly summarized as follows:
\begin{itemize}
\item The contribution from high-$z$ galaxies, with star formation efficiency set to $f_\star=0.005$ to match the LF from the deep HST Legacy surveys, cannot reproduce the levels of unresolved fluctuations in the near-IR for any reasonable extrapolation, producing merely $\sim 0.01-0.05$\nW\ flux at $z>8$. In fact, faint galaxies at intermediate-$z$ make a larger contribution but are themselves insufficient to account for the entire clustering signal \citep{Helgason12}. This conclusion is in agreement with previous studies \citep{Fernandez12,Cooray12a,Yue13a}.
\item A simple extrapolation of the faint-end of the LF suggests that high-$z$ galaxies will become significant once sources have been removed down to $m_{\rm AB}\gsim 32$. Using the CIB to probe these populations will therefore require  JWST \citep{Kashlinsky15} and/or cross-correlation techniques \citep{AtrioBarandela14}.
\item For more energetically efficient modes of star formation, we find that heavier IMFs still require high star formation efficiencies $f_\star>0.1$ which must either take place in small halos or end before $z\simeq 12$. The ionizing photons associated with such a population would have to be suppressed to avoid rapid reionization and to enhance the nebular emission contributing to the CIB. In other words, it would require very low escape fractions, $f_{\rm esc}<0.1\%$, in low-mass systems towards high redshifts \citep[see also ][]{Fernandez10}. This is contrary to what is found in theoretical studies \citep{FerraraLoeb13,Kuhlen12}.
\item The cumulative CIB contribution of the fossils of the first galaxies that contain old stellar populations radiating throughout cosmic history is comparable to that of their high-$z$ progenitors. Since many such systems will merge or become satellites of larger galaxies, their net contribution to the source-subtracted CIB is smaller and can be safely neglected as long as the contribution of the high-$z$ counterparts is also low.
\item For the CIB fluctuations to be produced by accreting black holes, one requires vigorous accretion rates in the early universe reaching $\rho_{\rm acc} \gsim 10^5M_\odot{\rm Mpc^{-3}}$ by $z=10$. This quantitative limit assumes a constant efficiency but is mostly independent of the accretion rate history and BH seed masses. The accretion must be very radiatively efficient at early times and then drop to inefficient growth throughout most of the remaining cosmic time with the resulting population largely unaccounted for in the SMBH census in local universe \citep{Vika09}. In addition, this population must be extremely gas-obscured in order to i) avoid reionizing the universe too rapidly, ii) to suppress the associated soft X-ray background, iii) to enhance the fraction of energy emitted in the UV as reprocessed nebular emission. \citet{Yue13b} have designed a novel model of DCBH formation that is broadly consistent with these requirements. % (see discussion below).
\item For the expected level of clustering of sources (both at low and high-$z$), the minimum soft CXB flux implied by the measured CIB$\times$CXB cross-correlation, $F_{\rm X} \simeq 7 \times 10^{-13} ~ {\rm erg~s^{-1}cm^{-2}deg^{-2}}$, is only marginally consistent with the most stringent CXB limits \citep{Moretti12}. At face value this means that either i) that high-$z$ BHs require the entire available unresolved CXB, ii) the clustering of the underlying sources is stronger (e.g. more biased) with $\Delta_{\rm cl}\gsim 0.2$ at sub-degree scales, or iii) the high current uncertainty of the CIB$\times$CXB level at large scales could eventually bring it into agreement with low-$z$ AGN and XRBs \citep{Helgason14}.

\end{itemize}

In this paper, we explored the physical requirements for stellar emission and accretion separately. It is however entirely possible, and in fact plausible, that the CIB and CXB are not produced by the same physical sources. The most intuitive scenario would have the CIB produced by stars, whereas accreting black holes sharing the same structures are responsible for the CIB$\times$CXB correlation. This would be a more natural way to obtain the $F_{\rm IR}/F_{\rm X}\sim 100$ ratio, which is unconventional for BHs only. The requirements for BHs to produce the CIB$\times$CXB component only, either stellar remnants or DCBHs, are much less energetically demanding than for the single-population explanation of the CIB and CIB$\times$CXB fluctuations. However, whereas rapid reionization is more easily avoided, one still requires the entire unresolved CXB available and is then left with the problem of explaining the CIB, the subject of this paper, by other means.

Our study of stellar sources isolates a narrow corner of parameters required for explaining the properties of the measured CIB fluctuations in terms of physics and evolution at high-$z$. The predominant contributions would then have to come from very massive stars ($\gsim 100M_\odot$), and accreting black holes, with non-instantaneous reionization in a highly clumped IGM. The area covered by this narrow corner may be widened by a factor of a few by relaxing assumptions of constancy of parameters such as $f_\star$ or $\epsilon$. Apart from nucleosynthesis and accretion, an additional emission component at high $z$ may come from stellar collisions in the dense stellar system phase which may result from stellar dynamical evolution of the early systems \citep[e.g. ][]{KashlinskyRees83}. The specific weight of this component is hard to quantify in a robust model-independent manner, but it may result in additional energy releases and ultimately formation of supermassive BHs \citep{BegelmanRees78}. It is important to emphasize that narrow allowed parameter-space is predicated on 1) the assumption of accurate estimate of the power contained in the primordial density spectrum at very small scales (comoving kpc), and 2) the assumption of its time-invariance after the first sources form. The first drives the collapse of the first halos with tiny modifications in the amount of power leading to exponentially different $f_{\rm col}(z)$ and, consequently, different, potentially larger CIB. The second possibility may arise from the inevitable and highly complex interaction between the condition of the density field which forms early collapsing objects and the energy releasing sources \citep[e.g. ][]{OstrikerCowie81,Rees85}. In this context, recent studies have found that high-$z$ galaxies may have had to form surprisingly early to account for the weakening of the exponential decline at the bright end of the LF with increasing redshift \citet{Finkelstein15}. This could be achieved by an unusually early collapse with respect to standard clustering theory, or an increased star formation efficiency making the stellar-to-halo mass ratio appear to rise with increasing redshift \citep{Steinhardt15}. In order to have a significant impact on the CIB however, these effect must also be present in lower mass systems which contribute the bulk of the CIB.

All this argues for empirically measuring the CIB fluctuations with required fidelity, establishing its nature and then constructing and identifying the implied physics models for the sources producing the signal with all its properties. This has been demonstrated in multi-wavelength cross-correlation studies that provide further constraints for theoretical models of both low-$z$ and high-$z$ sources \citep[e.g. ][]{Cappelluti13,Thacker14}.  Additional information on the physics and evolution at early times would come by looking for correlations between the CIB and 21 cm maps \citep{Fernandez14, Mao14}. {\it Euclid}'s large sky coverage and wavelengths make it particularly suitable for using its to-be-measured CIB fluctuations to isolate emissions from the epoch of re-ionization (Kashlinsky et al. 2015, submitted) and identifying the condition of the IGM by cross-correlating {\it Euclid}'s measured source-subtracted all-sky CIB and CMB maps \citep{AtrioBarandela14}. A specific experiment has been recently proposed with the {\it JWST}/NIRCAM to probe the Lyman-break of the CIB component and to tomographically reconstruct the emissions out to $z\gsim 30$ \citep{Kashlinsky15}; the Lyman tomography method proposed there was shown to already lead to interesting limits at $z\gsim 30$ from the available {\it Spitzer} CIB maps at 3.6 and 4.5\mic. The nature of the CIB fluctuations should be resolvable in the coming decade and the emissions produced during the epoch of reionization identified from its properties, shedding light on the physics and evolution at that time.

\section*{ACKNOWLEDGMENTS}

We thank B. Yue, A. Ferrara, N. Cappelluti and E. Komatsu for useful discussions. KH was supported by the European Union’s Seventh Framework Programme (FP7-PEOPLE-2013-IFF) under grant agreement number 628319-CIBorigins; and by NASA Headquarters under the NESSF Program Grant -- NNX11AO05H. MR acknowledges support from NSF CDI-typeII grant CMMI1125285 and the Theoretical and Computational Astrophysics Network (TCAN) grant AST1333514. AK acknowledges NASA's support for the Euclid LIBRAE project NNN12AA01C. VB was supported by NSF grant AST-1413501.

%\bibliographystyle{/Users/kari/Dropbox/work/tex/apj}
%\bibliography{/Users/kari/Dropbox/work/tex/myrefs}

\begin{thebibliography}{112}
\expandafter\ifx\csname natexlab\endcsname\relax\def\natexlab#1{#1}\fi

\bibitem[{{Ackermann et al.}(2012)}]{Ackermann12}
{Ackermann et al.} 2012, Science, 338, 1190

\bibitem[{{Arendt} {et~al.}(2010){Arendt}, {Kashlinsky}, {Moseley}, \&
  {Mather}}]{Arendt10}
{Arendt}, R.~G., {Kashlinsky}, A., {Moseley}, S.~H., \& {Mather}, J. 2010,
  \apjs, 186, 10

\bibitem[{{Ashby} {et~al.}(2013){Ashby}, {Willner}, {Fazio}, {Huang}, {Arendt},
  {Barmby}, {Barro}, {Bell}, {Bouwens}, {Cattaneo}, {Croton}, {Dav{\'e}},
  {Dunlop}, {Egami}, {Faber}, {Finlator}, {Grogin}, {Guhathakurta},
  {Hernquist}, {Hora}, {Illingworth}, {Kashlinsky}, {Koekemoer}, {Koo},
  {Labb{\'e}}, {Li}, {Lin}, {Moseley}, {Nandra}, {Newman}, {Noeske}, {Ouchi},
  {Peth}, {Rigopoulou}, {Robertson}, {Sarajedini}, {Simard}, {Smith}, {Wang},
  {Wechsler}, {Weiner}, {Wilson}, {Wuyts}, {Yamada}, \& {Yan}}]{Ashby13}
{Ashby}, M.~L.~N., {Willner}, S.~P., {Fazio}, G.~G., {Huang}, J.-S., {Arendt},
  R., {Barmby}, P., {Barro}, G., {Bell}, E.~F., {Bouwens}, R., {Cattaneo}, A.,
  {Croton}, D., {Dav{\'e}}, R., {Dunlop}, J.~S., {Egami}, E., {Faber}, S.,
  {Finlator}, K., {Grogin}, N.~A., {Guhathakurta}, P., {Hernquist}, L., {Hora},
  J.~L., {Illingworth}, G., {Kashlinsky}, A., {Koekemoer}, A.~M., {Koo}, D.~C.,
  {Labb{\'e}}, I., {Li}, Y., {Lin}, L., {Moseley}, H., {Nandra}, K., {Newman},
  J., {Noeske}, K., {Ouchi}, M., {Peth}, M., {Rigopoulou}, D., {Robertson}, B.,
  {Sarajedini}, V., {Simard}, L., {Smith}, H.~A., {Wang}, Z., {Wechsler}, R.,
  {Weiner}, B., {Wilson}, G., {Wuyts}, S., {Yamada}, T., \& {Yan}, H. 2013,
  \apj, 769, 80

\bibitem[{{Atrio-Barandela} \& {Kashlinsky}(2014)}]{AtrioBarandela14}
{Atrio-Barandela}, F. \& {Kashlinsky}, A. 2014, \apjl, 797, L26

\bibitem[{{Barkat} {et~al.}(1967){Barkat}, {Rakavy}, \& {Sack}}]{Barkat67}
{Barkat}, Z., {Rakavy}, G., \& {Sack}, N. 1967, Physical Review Letters, 18,
  379

\bibitem[{{Begelman} \& {Rees}(1978)}]{BegelmanRees78}
{Begelman}, M.~C. \& {Rees}, M.~J. 1978, \mnras, 185, 847

\bibitem[{{Behroozi} {et~al.}(2013){Behroozi}, {Wechsler}, \&
  {Conroy}}]{Behroozi13}
{Behroozi}, P.~S., {Wechsler}, R.~H., \& {Conroy}, C. 2013, \apj, 770, 57

\bibitem[{{Biteau} \& {Williams}(2015)}]{Biteau15}
{Biteau}, J. \& {Williams}, D.~A. 2015, ArXiv e-prints

\bibitem[{{Bouwens} {et~al.}(2007){Bouwens}, {Illingworth}, {Franx}, \&
  {Ford}}]{Bouwens07}
{Bouwens}, R.~J., {Illingworth}, G.~D., {Franx}, M., \& {Ford}, H. 2007, \apj,
  670, 928

\bibitem[{{Bouwens} {et~al.}(2014){Bouwens}, {Illingworth}, {Oesch}, {Trenti},
  {Labbe'}, {Bradley}, {Carollo}, {van Dokkum}, {Gonzalez}, {Holwerda},
  {Franx}, {Spitler}, {Smit}, \& {Magee}}]{Bouwens14}
{Bouwens}, R.~J., {Illingworth}, G.~D., {Oesch}, P.~A., {Trenti}, M., {Labbe'},
  I., {Bradley}, L., {Carollo}, M., {van Dokkum}, P.~G., {Gonzalez}, V.,
  {Holwerda}, B., {Franx}, M., {Spitler}, L., {Smit}, R., \& {Magee}, D. 2014,
  ArXiv e-prints

\bibitem[{{Bovill} \& {Ricotti}(2009)}]{BovillRicotti09}
{Bovill}, M.~S. \& {Ricotti}, M. 2009, \apj, 693, 1859

\bibitem[{{Bovill} \& {Ricotti}(2011{\natexlab{a}})}]{Bovill11a}
---. 2011{\natexlab{a}}, \apj, 741, 17

\bibitem[{{Bovill} \& {Ricotti}(2011{\natexlab{b}})}]{Bovill11b}
---. 2011{\natexlab{b}}, \apj, 741, 18

\bibitem[{{Bromm}(2013)}]{Bromm13}
{Bromm}, V. 2013, Reports on Progress in Physics, 76, 112901

\bibitem[{{Cappelluti} {et~al.}(2013){Cappelluti}, {Kashlinsky}, {Arendt},
  {Comastri}, {Fazio}, {Finoguenov}, {Hasinger}, {Mather}, {Miyaji}, \&
  {Moseley}}]{Cappelluti13}
{Cappelluti}, N., {Kashlinsky}, A., {Arendt}, R.~G., {Comastri}, A., {Fazio},
  G.~G., {Finoguenov}, A., {Hasinger}, G., {Mather}, J.~C., {Miyaji}, T., \&
  {Moseley}, S.~H. 2013, \apj, 769, 68

\bibitem[{{Comastri} {et~al.}(2015){Comastri}, {Gilli}, {Marconi}, {Risaliti},
  \& {Salvati}}]{Comastri15}
{Comastri}, A., {Gilli}, R., {Marconi}, A., {Risaliti}, G., \& {Salvati}, M.
  2015, \aap, 574, L10

\bibitem[{{Cooray} {et~al.}(2004){Cooray}, {Bock}, {Keatin}, {Lange}, \&
  {Matsumoto}}]{Cooray04}
{Cooray}, A., {Bock}, J.~J., {Keatin}, B., {Lange}, A.~E., \& {Matsumoto}, T.
  2004, \apj, 606, 611

\bibitem[{{Cooray} {et~al.}(2012{\natexlab{a}}){Cooray}, {Gong}, {Smidt}, \&
  {Santos}}]{Cooray12a}
{Cooray}, A., {Gong}, Y., {Smidt}, J., \& {Santos}, M.~G. 2012{\natexlab{a}},
  \apj, 756, 92

\bibitem[{{Cooray} \& {Sheth}(2002)}]{CooraySheth02}
{Cooray}, A. \& {Sheth}, R. 2002, \physrep, 372, 1

\bibitem[{{Cooray} {et~al.}(2012{\natexlab{b}}){Cooray}, {Smidt}, {de
  Bernardis}, {Gong}, {Stern}, {Ashby}, {Eisenhardt}, {Frazer}, {Gonzalez},
  {Kochanek}, {Kozlowski}, \& {Wright}}]{Cooray12b}
{Cooray}, A., {Smidt}, J., {de Bernardis}, F., {Gong}, Y., {Stern}, D.,
  {Ashby}, M.~L.~N., {Eisenhardt}, P., {Frazer}, C.~C., {Gonzalez}, A.~H.,
  {Kochanek}, C.~S., {Kozlowski}, S., \& {Wright}, E.~L. 2012{\natexlab{b}},
  \nat, 490, 514

\bibitem[{{Cooray} \& {Yoshida}(2004)}]{CoorayYoshida04}
{Cooray}, A. \& {Yoshida}, N. 2004, \mnras, 351, L71

\bibitem[{{Dwek} \& {Arendt}(1998)}]{DwekArendt98}
{Dwek}, E. \& {Arendt}, R.~G. 1998, \apjl, 508, L9

\bibitem[{{Eisenstein} \& {Hu}(1998)}]{EisensteinHu98}
{Eisenstein}, D.~J. \& {Hu}, W. 1998, \apj, 496, 605

\bibitem[{{Ellis} {et~al.}(2013){Ellis}, {McLure}, {Dunlop}, {Robertson},
  {Ono}, {Schenker}, {Koekemoer}, {Bowler}, {Ouchi}, {Rogers}, {Curtis-Lake},
  {Schneider}, {Charlot}, {Stark}, {Furlanetto}, \& {Cirasuolo}}]{Ellis13}
{Ellis}, R.~S., {McLure}, R.~J., {Dunlop}, J.~S., {Robertson}, B.~E., {Ono},
  Y., {Schenker}, M.~A., {Koekemoer}, A., {Bowler}, R.~A.~A., {Ouchi}, M.,
  {Rogers}, A.~B., {Curtis-Lake}, E., {Schneider}, E., {Charlot}, S., {Stark},
  D.~P., {Furlanetto}, S.~R., \& {Cirasuolo}, M. 2013, \apjl, 763, L7

\bibitem[{{Fazio} {et~al.}(2004){Fazio}, {Ashby}, {Barmby}, {Hora}, {Huang},
  {Pahre}, {Wang}, {Willner}, {Arendt}, {Moseley}, {Brodwin}, {Eisenhardt},
  {Stern}, {Tollestrup}, \& {Wright}}]{Fazio04}
{Fazio}, G.~G., {Ashby}, M.~L.~N., {Barmby}, P., {Hora}, J.~L., {Huang}, J.-S.,
  {Pahre}, M.~A., {Wang}, Z., {Willner}, S.~P., {Arendt}, R.~G., {Moseley},
  S.~H., {Brodwin}, M., {Eisenhardt}, P., {Stern}, D., {Tollestrup}, E.~V., \&
  {Wright}, E.~L. 2004, \apjs, 154, 39

\bibitem[{{Fernandez} {et~al.}(2012){Fernandez}, {Iliev}, {Komatsu}, \&
  {Shapiro}}]{Fernandez12}
{Fernandez}, E.~R., {Iliev}, I.~T., {Komatsu}, E., \& {Shapiro}, P.~R. 2012,
  \apj, 750, 20

\bibitem[{{Fernandez} \& {Komatsu}(2006)}]{Fernandez06}
{Fernandez}, E.~R. \& {Komatsu}, E. 2006, \apj, 646, 703

\bibitem[{{Fernandez} {et~al.}(2010){Fernandez}, {Komatsu}, {Iliev}, \&
  {Shapiro}}]{Fernandez10}
{Fernandez}, E.~R., {Komatsu}, E., {Iliev}, I.~T., \& {Shapiro}, P.~R. 2010,
  \apj, 710, 1089

\bibitem[{{Fernandez} {et~al.}(2014){Fernandez}, {Zaroubi}, {Iliev}, {Mellema},
  \& {Jeli{\'c}}}]{Fernandez14}
{Fernandez}, E.~R., {Zaroubi}, S., {Iliev}, I.~T., {Mellema}, G., \&
  {Jeli{\'c}}, V. 2014, \mnras, 440, 298

\bibitem[{{Ferrara} \& {Loeb}(2013)}]{FerraraLoeb13}
{Ferrara}, A. \& {Loeb}, A. 2013, \mnras, 431, 2826

\bibitem[{{Finkelstein} {et~al.}(2014){Finkelstein}, {Ryan}, {Papovich},
  {Dickinson}, {Song}, {Somerville}, {Ferguson}, {Salmon}, {Giavalisco},
  {Koekemoer}, {Ashby}, {Behroozi}, {Castellano}, {Dunlop}, {Faber}, {Fazio},
  {Fontana}, {Grogin}, {Hathi}, {Jaacks}, {Kocevski}, {Livermore}, {McLure},
  {Merlin}, {Mobasher}, {Newman}, {Rafelski}, {Tilvi}, \&
  {Willner}}]{Finkelstein14}
{Finkelstein}, S.~L., {Ryan}, Jr., R.~E., {Papovich}, C., {Dickinson}, M.,
  {Song}, M., {Somerville}, R., {Ferguson}, H.~C., {Salmon}, B., {Giavalisco},
  M., {Koekemoer}, A.~M., {Ashby}, M.~L.~N., {Behroozi}, P., {Castellano}, M.,
  {Dunlop}, J.~S., {Faber}, S.~M., {Fazio}, G.~G., {Fontana}, A., {Grogin},
  N.~A., {Hathi}, N., {Jaacks}, J., {Kocevski}, D.~D., {Livermore}, R.,
  {McLure}, R.~J., {Merlin}, E., {Mobasher}, B., {Newman}, J.~A., {Rafelski},
  M., {Tilvi}, V., \& {Willner}, S.~P. 2014, ArXiv e-prints

\bibitem[{{Finkelstein} {et~al.}(2015){Finkelstein}, {Song}, {Behroozi},
  {Somerville}, {Papovich}, {Milosavljevic}, {Dekel}, {Narayanan}, {Ashby},
  {Cooray}, {Fazio}, {Ferguson}, {Koekemoer}, {Salmon}, \&
  {Willner}}]{Finkelstein15}
{Finkelstein}, S.~L., {Song}, M., {Behroozi}, P., {Somerville}, R.~S.,
  {Papovich}, C., {Milosavljevic}, M., {Dekel}, A., {Narayanan}, D., {Ashby},
  M.~L.~N., {Cooray}, A., {Fazio}, G.~G., {Ferguson}, H.~C., {Koekemoer},
  A.~M., {Salmon}, B.~W., \& {Willner}, S.~P. 2015, ArXiv e-prints

\bibitem[{{Furlanetto} \& {Loeb}(2005)}]{FurlanettoLoeb05}
{Furlanetto}, S.~R. \& {Loeb}, A. 2005, \apj, 634, 1

\bibitem[{{Greif} \& {Bromm}(2006)}]{GreifBromm06}
{Greif}, T.~H. \& {Bromm}, V. 2006, \mnras, 373, 128

\bibitem[{{Helgason} {et~al.}(2014){Helgason}, {Cappelluti}, {Hasinger},
  {Kashlinsky}, \& {Ricotti}}]{Helgason14}
{Helgason}, K., {Cappelluti}, N., {Hasinger}, G., {Kashlinsky}, A., \&
  {Ricotti}, M. 2014, \apj, 785, 38

\bibitem[{{Helgason} {et~al.}(2012){Helgason}, {Ricotti}, \&
  {Kashlinsky}}]{Helgason12}
{Helgason}, K., {Ricotti}, M., \& {Kashlinsky}, A. 2012, \apj, 752, 113

\bibitem[{{H.E.S.S.~Collaboration}(2013)}]{HESS13}
{H.E.S.S.~Collaboration}. 2013, \aap, 550, A4

\bibitem[{{Hickox} \& {Markevitch}(2006)}]{Hickox06}
{Hickox}, R.~C. \& {Markevitch}, M. 2006, \apj, 645, 95

\bibitem[{{Hinshaw} {et~al.}(2013){Hinshaw}, {Larson}, {Komatsu}, {Spergel},
  {Bennett}, {Dunkley}, {Nolta}, {Halpern}, {Hill}, {Odegard}, {Page}, {Smith},
  {Weiland}, {Gold}, {Jarosik}, {Kogut}, {Limon}, {Meyer}, {Tucker}, {Wollack},
  \& {Wright}}]{Hinshaw13}
{Hinshaw}, G., {Larson}, D., {Komatsu}, E., {Spergel}, D.~N., {Bennett}, C.~L.,
  {Dunkley}, J., {Nolta}, M.~R., {Halpern}, M., {Hill}, R.~S., {Odegard}, N.,
  {Page}, L., {Smith}, K.~M., {Weiland}, J.~L., {Gold}, B., {Jarosik}, N.,
  {Kogut}, A., {Limon}, M., {Meyer}, S.~S., {Tucker}, G.~S., {Wollack}, E., \&
  {Wright}, E.~L. 2013, \apjs, 208, 19

\bibitem[{{Hummer}(1994)}]{Hummer94}
{Hummer}, D.~G. 1994, \mnras, 268, 109

\bibitem[{{Iwamoto} {et~al.}(1998){Iwamoto}, {Mazzali}, {Nomoto}, {Umeda},
  {Nakamura}, {Patat}, {Danziger}, {Young}, {Suzuki}, {Shigeyama},
  {Augusteijn}, {Doublier}, {Gonzalez}, {Boehnhardt}, {Brewer}, {Hainaut},
  {Lidman}, {Leibundgut}, {Cappellaro}, {Turatto}, {Galama}, {Vreeswijk},
  {Kouveliotou}, {van Paradijs}, {Pian}, {Palazzi}, \& {Frontera}}]{Iwamoto98}
{Iwamoto}, K., {Mazzali}, P.~A., {Nomoto}, K., {Umeda}, H., {Nakamura}, T.,
  {Patat}, F., {Danziger}, I.~J., {Young}, T.~R., {Suzuki}, T., {Shigeyama},
  T., {Augusteijn}, T., {Doublier}, V., {Gonzalez}, J.-F., {Boehnhardt}, H.,
  {Brewer}, J., {Hainaut}, O.~R., {Lidman}, C., {Leibundgut}, B., {Cappellaro},
  E., {Turatto}, M., {Galama}, T.~J., {Vreeswijk}, P.~M., {Kouveliotou}, C.,
  {van Paradijs}, J., {Pian}, E., {Palazzi}, E., \& {Frontera}, F. 1998, \nat,
  395, 672

\bibitem[{{Jeon} {et~al.}(2014){Jeon}, {Pawlik}, {Bromm}, \&
  {Milosavljevi{\'c}}}]{Jeon14}
{Jeon}, M., {Pawlik}, A.~H., {Bromm}, V., \& {Milosavljevi{\'c}}, M. 2014,
  \mnras, 444, 3288

\bibitem[{{Johnson} {et~al.}(2009){Johnson}, {Greif}, {Bromm}, {Klessen}, \&
  {Ippolito}}]{Johnson09}
{Johnson}, J.~L., {Greif}, T.~H., {Bromm}, V., {Klessen}, R.~S., \& {Ippolito},
  J. 2009, \mnras, 399, 37

\bibitem[{{Kashlinsky}(2005)}]{Kashreview}
{Kashlinsky}, A. 2005, \physrep, 409, 361

\bibitem[{{Kashlinsky} {et~al.}(2004){Kashlinsky}, {Arendt}, {Gardner},
  {Mather}, \& {Moseley}}]{Kashlinsky04}
{Kashlinsky}, A., {Arendt}, R., {Gardner}, J.~P., {Mather}, J.~C., \&
  {Moseley}, S.~H. 2004, \apj, 608, 1

\bibitem[{{Kashlinsky} {et~al.}(2012){Kashlinsky}, {Arendt}, {Ashby}, {Fazio},
  {Mather}, \& {Moseley}}]{Kashlinsky12}
{Kashlinsky}, A., {Arendt}, R.~G., {Ashby}, M.~L.~N., {Fazio}, G.~G., {Mather},
  J., \& {Moseley}, S.~H. 2012, \apj, 753, 63

\bibitem[{{Kashlinsky} {et~al.}(2005){Kashlinsky}, {Arendt}, {Mather}, \&
  {Moseley}}]{KAMM1}
{Kashlinsky}, A., {Arendt}, R.~G., {Mather}, J., \& {Moseley}, S.~H. 2005,
  \nat, 438, 45

\bibitem[{{Kashlinsky} {et~al.}(2007{\natexlab{a}}){Kashlinsky}, {Arendt},
  {Mather}, \& {Moseley}}]{KAMM4}
---. 2007{\natexlab{a}}, \apjl, 666, L1

\bibitem[{{Kashlinsky} {et~al.}(2007{\natexlab{b}}){Kashlinsky}, {Arendt},
  {Mather}, \& {Moseley}}]{KAMM2}
---. 2007{\natexlab{b}}, \apjl, 654, L5

\bibitem[{{Kashlinsky} {et~al.}(2007{\natexlab{c}}){Kashlinsky}, {Arendt},
  {Mather}, \& {Moseley}}]{KAMM3}
---. 2007{\natexlab{c}}, \apjl, 654, L1

\bibitem[{{Kashlinsky} {et~al.}(2015){Kashlinsky}, {Mather}, {Helgason},
  {Arendt}, {Bromm}, \& {Moseley}}]{Kashlinsky15}
{Kashlinsky}, A., {Mather}, J.~C., {Helgason}, K., {Arendt}, R.~G., {Bromm},
  V., \& {Moseley}, S.~H. 2015, \apj, 804, 99

\bibitem[{{Kashlinsky} {et~al.}(2002){Kashlinsky}, {Odenwald}, {Mather},
  {Skrutskie}, \& {Cutri}}]{Kashlinsky02}
{Kashlinsky}, A., {Odenwald}, S., {Mather}, J., {Skrutskie}, M.~F., \& {Cutri},
  R.~M. 2002, \apjl, 579, L53

\bibitem[{{Kashlinsky} \& {Rees}(1983)}]{KashlinskyRees83}
{Kashlinsky}, A. \& {Rees}, M.~J. 1983, \mnras, 205, 955

\bibitem[{{Keenan} {et~al.}(2010){Keenan}, {Trouille}, {Barger}, {Cowie}, \&
  {Wang}}]{Keenan10a}
{Keenan}, R.~C., {Trouille}, L., {Barger}, A.~J., {Cowie}, L.~L., \& {Wang},
  W.-H. 2010, \apjs, 186, 94

\bibitem[{{Kuhlen} \& {Faucher-Gigu{\`e}re}(2012)}]{Kuhlen12}
{Kuhlen}, M. \& {Faucher-Gigu{\`e}re}, C.-A. 2012, \mnras, 423, 862

\bibitem[{{Leitherer} {et~al.}(1999){Leitherer}, {Schaerer}, {Goldader},
  {Gonz{\'a}lez Delgado}, {Robert}, {Kune}, {de Mello}, {Devost}, \&
  {Heckman}}]{Leitherer99}
{Leitherer}, C., {Schaerer}, D., {Goldader}, J.~D., {Gonz{\'a}lez Delgado},
  R.~M., {Robert}, C., {Kune}, D.~F., {de Mello}, D.~F., {Devost}, D., \&
  {Heckman}, T.~M. 1999, \apjs, 123, 3

\bibitem[{{Limber}(1953)}]{Limber53}
{Limber}, D.~N. 1953, \apj, 117, 134

\bibitem[{{Madau} \& {Pozzetti}(2000)}]{MadauPozzetti00}
{Madau}, P. \& {Pozzetti}, L. 2000, \mnras, 312, L9

\bibitem[{{Madau} {et~al.}(2014){Madau}, {Weisz}, \& {Conroy}}]{Madau14}
{Madau}, P., {Weisz}, D.~R., \& {Conroy}, C. 2014, \apjl, 790, L17

\bibitem[{{Mao}(2014)}]{Mao14}
{Mao}, X.-C. 2014, \apj, 790, 148

\bibitem[{{Marconi} {et~al.}(2004){Marconi}, {Risaliti}, {Gilli}, {Hunt},
  {Maiolino}, \& {Salvati}}]{Marconi04}
{Marconi}, A., {Risaliti}, G., {Gilli}, R., {Hunt}, L.~K., {Maiolino}, R., \&
  {Salvati}, M. 2004, \mnras, 351, 169

\bibitem[{{Matsumoto} {et~al.}(2015){Matsumoto}, {Kim}, {Pyo}, \&
  {Tsumura}}]{Matsumoto15}
{Matsumoto}, T., {Kim}, M.~G., {Pyo}, J., \& {Tsumura}, K. 2015, ArXiv e-prints

\bibitem[{{Matsumoto} {et~al.}(2005){Matsumoto}, {Matsuura}, {Murakami},
  {Tanaka}, {Freund}, {Lim}, {Cohen}, {Kawada}, \& {Noda}}]{Matsumoto05}
{Matsumoto}, T., {Matsuura}, S., {Murakami}, H., {Tanaka}, M., {Freund}, M.,
  {Lim}, M., {Cohen}, M., {Kawada}, M., \& {Noda}, M. 2005, \apj, 626, 31

\bibitem[{{Matsumoto} {et~al.}(2011){Matsumoto}, {Seo}, {Jeong}, {Lee},
  {Matsuura}, {Matsuhara}, {Oyabu}, {Pyo}, \& {Wada}}]{Matsumoto11}
{Matsumoto}, T., {Seo}, H.~J., {Jeong}, W.-S., {Lee}, H.~M., {Matsuura}, S.,
  {Matsuhara}, H., {Oyabu}, S., {Pyo}, J., \& {Wada}, T. 2011, \apj, 742, 124

\bibitem[{{Mazin} \& {Raue}(2007)}]{MazinRaue07}
{Mazin}, D. \& {Raue}, M. 2007, \aap, 471, 439

\bibitem[{{McLure} {et~al.}(2009){McLure}, {Cirasuolo}, {Dunlop}, {Foucaud}, \&
  {Almaini}}]{McLure09}
{McLure}, R.~J., {Cirasuolo}, M., {Dunlop}, J.~S., {Foucaud}, S., \& {Almaini},
  O. 2009, \mnras, 395, 2196

\bibitem[{{Meyer} {et~al.}(2012){Meyer}, {Raue}, {Mazin}, \& {Horns}}]{Meyer12}
{Meyer}, M., {Raue}, M., {Mazin}, D., \& {Horns}, D. 2012, \aap, 542, A59

\bibitem[{{Mitra} {et~al.}(2013){Mitra}, {Ferrara}, \& {Choudhury}}]{Mitra13}
{Mitra}, S., {Ferrara}, A., \& {Choudhury}, T.~R. 2013, \mnras, 428, L1

\bibitem[{{Moretti} {et~al.}(2012){Moretti}, {Vattakunnel}, {Tozzi},
  {Salvaterra}, {Severgnini}, {Fugazza}, {Haardt}, \& {Gilli}}]{Moretti12}
{Moretti}, A., {Vattakunnel}, S., {Tozzi}, P., {Salvaterra}, R., {Severgnini},
  P., {Fugazza}, D., {Haardt}, F., \& {Gilli}, R. 2012, \aap, 548, A87

\bibitem[{{Oke} \& {Gunn}(1983)}]{OkeGunn83}
{Oke}, J.~B. \& {Gunn}, J.~E. 1983, \apj, 266, 713

\bibitem[{{O'Shea} {et~al.}(2015){O'Shea}, {Wise}, {Xu}, \& {Norman}}]{OShea15}
{O'Shea}, B.~W., {Wise}, J.~H., {Xu}, H., \& {Norman}, M.~L. 2015, ArXiv
  e-prints

\bibitem[{{Ostriker} \& {Cowie}(1981)}]{OstrikerCowie81}
{Ostriker}, J.~P. \& {Cowie}, L.~L. 1981, \apjl, 243, L127

\bibitem[{{Pacucci} \& {Ferrara}(2015)}]{Pacucci15}
{Pacucci}, F. \& {Ferrara}, A. 2015, \mnras, 448, 104

\bibitem[{{Pawlik} {et~al.}(2015){Pawlik}, {Schaye}, \& {Dalla
  Vecchia}}]{Pawlik14}
{Pawlik}, A.~H., {Schaye}, J., \& {Dalla Vecchia}, C. 2015, ArXiv e-prints

\bibitem[{{Planck Collaboration} {et~al.}(2013){Planck Collaboration}, {Ade},
  {Aghanim}, {Armitage-Caplan}, {Arnaud}, {Ashdown}, {Atrio-Barandela},
  {Aumont}, {Baccigalupi}, {Banday}, \& et~al.}]{PlanckCosmology}
{Planck Collaboration}, {Ade}, P.~A.~R., {Aghanim}, N., {Armitage-Caplan}, C.,
  {Arnaud}, M., {Ashdown}, M., {Atrio-Barandela}, F., {Aumont}, J.,
  {Baccigalupi}, C., {Banday}, A.~J., \& et~al. 2013, ArXiv e-prints

\bibitem[{{Planck Collaboration} {et~al.}(2015){Planck Collaboration}, {Ade},
  {Aghanim}, {Arnaud}, {Ashdown}, {Aumont}, {Baccigalupi}, {Banday},
  {Barreiro}, {Bartlett}, \& et~al.}]{Planck15}
{Planck Collaboration}, {Ade}, P.~A.~R., {Aghanim}, N., {Arnaud}, M.,
  {Ashdown}, M., {Aumont}, J., {Baccigalupi}, C., {Banday}, A.~J., {Barreiro},
  R.~B., {Bartlett}, J.~G., \& et~al. 2015, ArXiv e-prints

\bibitem[{{Press} \& {Schechter}(1974)}]{PressSchechter74}
{Press}, W.~H. \& {Schechter}, P. 1974, \apj, 187, 425

\bibitem[{{Rees}(1985)}]{Rees85}
{Rees}, M.~J. 1985, \mnras, 213, 75P

\bibitem[{{Ricotti} \& {Gnedin}(2005)}]{RicottiGnedin05}
{Ricotti}, M. \& {Gnedin}, N.~Y. 2005, \apj, 629, 259

\bibitem[{{Ricotti} {et~al.}(2002{\natexlab{a}}){Ricotti}, {Gnedin}, \&
  {Shull}}]{Ricotti02a}
{Ricotti}, M., {Gnedin}, N.~Y., \& {Shull}, J.~M. 2002{\natexlab{a}}, \apj,
  575, 33

\bibitem[{{Ricotti} {et~al.}(2002{\natexlab{b}}){Ricotti}, {Gnedin}, \&
  {Shull}}]{Ricotti02b}
---. 2002{\natexlab{b}}, \apj, 575, 49

\bibitem[{{Ricotti} {et~al.}(2008){Ricotti}, {Gnedin}, \& {Shull}}]{Ricotti08}
---. 2008, \apj, 685, 21

\bibitem[{{Ricotti} {et~al.}(2005){Ricotti}, {Ostriker}, \&
  {Gnedin}}]{Ricotti05a}
{Ricotti}, M., {Ostriker}, J.~P., \& {Gnedin}, N.~Y. 2005, \mnras, 357, 207

\bibitem[{{Ricotti} \& {Shull}(2000)}]{RicottiShull00}
{Ricotti}, M. \& {Shull}, J.~M. 2000, \apj, 542, 548

\bibitem[{{Robertson} {et~al.}(2015){Robertson}, {Ellis}, {Furlanetto}, \&
  {Dunlop}}]{Robertson15}
{Robertson}, B.~E., {Ellis}, R.~S., {Furlanetto}, S.~R., \& {Dunlop}, J.~S.
  2015, \apjl, 802, L19

\bibitem[{{Salvaterra} \& {Ferrara}(2006)}]{Salvaterra06b}
{Salvaterra}, R. \& {Ferrara}, A. 2006, \mnras, 367, L11

\bibitem[{{Salvaterra} {et~al.}(2012){Salvaterra}, {Haardt}, {Volonteri}, \&
  {Moretti}}]{Salvaterra12}
{Salvaterra}, R., {Haardt}, F., {Volonteri}, M., \& {Moretti}, A. 2012, \aap,
  545, L6

\bibitem[{{Salvaterra} {et~al.}(2006){Salvaterra}, {Magliocchetti}, {Ferrara},
  \& {Schneider}}]{Salvaterra06a}
{Salvaterra}, R., {Magliocchetti}, M., {Ferrara}, A., \& {Schneider}, R. 2006,
  \mnras, 368, L6

\bibitem[{{Santos} {et~al.}(2002){Santos}, {Bromm}, \&
  {Kamionkowski}}]{Santos02}
{Santos}, M.~R., {Bromm}, V., \& {Kamionkowski}, M. 2002, \mnras, 336, 1082

\bibitem[{{Schaerer}(2002)}]{Schaerer02}
{Schaerer}, D. 2002, \aap, 382, 28

\bibitem[{{Seljak}(2000)}]{Seljak00}
{Seljak}, U. 2000, \mnras, 318, 203

\bibitem[{{Seo} {et~al.}(2015){Seo}, {Lee}, {Matsumoto}, {Jeong}, {Lee}, \&
  {Pyo}}]{Seo15}
{Seo}, H.~J., {Lee}, H.~M., {Matsumoto}, T., {Jeong}, W.-S., {Lee}, M.~G., \&
  {Pyo}, J. 2015, ArXiv e-prints

\bibitem[{{Shankar} {et~al.}(2004){Shankar}, {Salucci}, {Granato}, {De Zotti},
  \& {Danese}}]{Shankar04}
{Shankar}, F., {Salucci}, P., {Granato}, G.~L., {De Zotti}, G., \& {Danese}, L.
  2004, \mnras, 354, 1020

\bibitem[{{Sheth} {et~al.}(2001){Sheth}, {Mo}, \& {Tormen}}]{ShethTormen01}
{Sheth}, R.~K., {Mo}, H.~J., \& {Tormen}, G. 2001, \mnras, 323, 1

\bibitem[{{Soltan}(1982)}]{Soltan82}
{Soltan}, A. 1982, \mnras, 200, 115

\bibitem[{{Steinhardt} {et~al.}(2015){Steinhardt}, {Capak}, {Masters}, \&
  {Speagle}}]{Steinhardt15}
{Steinhardt}, C.~L., {Capak}, P., {Masters}, D., \& {Speagle}, J.~S. 2015,
  ArXiv e-prints

\bibitem[{{Tanaka} {et~al.}(2012){Tanaka}, {Perna}, \& {Haiman}}]{Tanaka12}
{Tanaka}, T., {Perna}, R., \& {Haiman}, Z. 2012, \mnras, 425, 2974

\bibitem[{{Tanaka} \& {Li}(2014)}]{Tanaka14}
{Tanaka}, T.~L. \& {Li}, M. 2014, \mnras, 439, 1092

\bibitem[{{Thacker} {et~al.}(2014){Thacker}, {Gong}, {Cooray}, {De Bernardis},
  {Smidt}, \& {Mitchell-Wynne}}]{Thacker14}
{Thacker}, C., {Gong}, Y., {Cooray}, A., {De Bernardis}, F., {Smidt}, J., \&
  {Mitchell-Wynne}, K. 2014, ArXiv e-prints

\bibitem[{{Thompson} {et~al.}(2007{\natexlab{a}}){Thompson}, {Eisenstein},
  {Fan}, {Rieke}, \& {Kennicutt}}]{Thompson07a}
{Thompson}, R.~I., {Eisenstein}, D., {Fan}, X., {Rieke}, M., \& {Kennicutt},
  R.~C. 2007{\natexlab{a}}, \apj, 657, 669

\bibitem[{{Thompson} {et~al.}(2007{\natexlab{b}}){Thompson}, {Eisenstein},
  {Fan}, {Rieke}, \& {Kennicutt}}]{Thompson07b}
---. 2007{\natexlab{b}}, \apj, 666, 658

\bibitem[{{Tinker} {et~al.}(2008){Tinker}, {Kravtsov}, {Klypin}, {Abazajian},
  {Warren}, {Yepes}, {Gottl{\"o}ber}, \& {Holz}}]{Tinker08}
{Tinker}, J., {Kravtsov}, A.~V., {Klypin}, A., {Abazajian}, K., {Warren}, M.,
  {Yepes}, G., {Gottl{\"o}ber}, S., \& {Holz}, D.~E. 2008, \apj, 688, 709

\bibitem[{{Treister} {et~al.}(2013){Treister}, {Schawinski}, {Volonteri}, \&
  {Natarajan}}]{Treister13}
{Treister}, E., {Schawinski}, K., {Volonteri}, M., \& {Natarajan}, P. 2013,
  \apj, 778, 130

\bibitem[{{Tsumura} {et~al.}(2013){Tsumura}, {Matsumoto}, {Matsuura}, {Sakon},
  \& {Wada}}]{Tsumura13}
{Tsumura}, K., {Matsumoto}, T., {Matsuura}, S., {Sakon}, I., \& {Wada}, T.
  2013, ArXiv e-prints

\bibitem[{{Vika} {et~al.}(2009){Vika}, {Driver}, {Graham}, \& {Liske}}]{Vika09}
{Vika}, M., {Driver}, S.~P., {Graham}, A.~W., \& {Liske}, J. 2009, \mnras, 400,
  1451

\bibitem[{{Volonteri} \& {Bellovary}(2012)}]{VolonteriBellovary12}
{Volonteri}, M. \& {Bellovary}, J. 2012, Reports on Progress in Physics, 75,
  124901

\bibitem[{{Weisz} {et~al.}(2014){Weisz}, {Dolphin}, {Skillman}, {Holtzman},
  {Gilbert}, {Dalcanton}, \& {Williams}}]{Weisz14}
{Weisz}, D.~R., {Dolphin}, A.~E., {Skillman}, E.~D., {Holtzman}, J., {Gilbert},
  K.~M., {Dalcanton}, J.~J., \& {Williams}, B.~F. 2014, \apj, 789, 148

\bibitem[{{Yue} {et~al.}(2013{\natexlab{a}}){Yue}, {Ferrara}, {Salvaterra}, \&
  {Chen}}]{Yue13a}
{Yue}, B., {Ferrara}, A., {Salvaterra}, R., \& {Chen}, X. 2013{\natexlab{a}},
  \mnras, 431, 383

\bibitem[{{Yue} {et~al.}(2013{\natexlab{b}}){Yue}, {Ferrara}, {Salvaterra},
  {Xu}, \& {Chen}}]{Yue13b}
{Yue}, B., {Ferrara}, A., {Salvaterra}, R., {Xu}, Y., \& {Chen}, X.
  2013{\natexlab{b}}, \mnras, 433, 1556

\bibitem[{{Yue} {et~al.}(2014){Yue}, {Ferrara}, {Salvaterra}, {Xu}, \&
  {Chen}}]{Yue14}
---. 2014, \mnras, 440, 1263

\bibitem[{{Zackrisson} {et~al.}(2011){Zackrisson}, {Rydberg}, {Schaerer},
  {{\"O}stlin}, \& {Tuli}}]{Zackrisson11}
{Zackrisson}, E., {Rydberg}, C.-E., {Schaerer}, D., {{\"O}stlin}, G., \&
  {Tuli}, M. 2011, \apj, 740, 13

\bibitem[{{Zemcov} {et~al.}(2014){Zemcov}, {Smidt}, {Arai}, {Bock}, {Cooray},
  {Gong}, {Kim}, {Korngut}, {Lam}, {Lee}, {Matsumoto}, {Matsuura}, {Nam},
  {Roudier}, {Tsumura}, \& {Wada}}]{Zemcov14}
{Zemcov}, M., {Smidt}, J., {Arai}, T., {Bock}, J., {Cooray}, A., {Gong}, Y.,
  {Kim}, M.~G., {Korngut}, P., {Lam}, A., {Lee}, D.~H., {Matsumoto}, T.,
  {Matsuura}, S., {Nam}, U.~W., {Roudier}, G., {Tsumura}, K., \& {Wada}, T.
  2014, Science, 346, 732

\end{thebibliography}

\end{document}